\documentstyle[12pt,equ,epsf,floats]{article}
\begin{document}
\title{INHOMOGENEOUS SYSTEMS AND THEIR RECTIFICATION PROPERTIES}

\author{A. M. Jayannavar \\ 
Institute of Physics, Sachivalaya Marg, Bhubaneswar-751005, INDIA.}
\date{ }

\maketitle

\begin{abstract}
We explore the possibility of obtaining unidirectional current in a
symmetric (periodic) potential system 
without the application of any obvious (apparent) externally applied bias.
There are many physical models proposed to accomplish this nonequilibrium effect.
In the present work we consider inhomogeneous systems 
so that the friction coefficient and/or temperature
could vary in space. We find out a model with minimal conditions that the inhomogeneous system
assisted by fluctuating forces must satisfy, in order to obtain unidirectional current. In the
process we discuss about thermal and frictional ratchets that are of current interest.
We argue that different models of frictional ratchets work under the same basic
principle of alteration of relative stability of otherwise
locally stable states in the presence of temperature inhomogeneity.
We also discuss in detail the nature of currents in rocked frictional ratchets. In particular we analyse a novel phenomenon of multiple current reversals and  the efficiency of the energy transduction in these systems.
 
\end{abstract}
\eject
\section{Introduction}

\par The study of interplay of noise and nonlinear dynamics presents many challenges in systems under non equilibrium conditions.
These systems  exhibit wide variety of novel physical outcomes(simple to complex).
 A nonequilibrium system is one in which there is a net
energy flow from external sources. Many different models employing  stochastic processes
in physics, chemistry, engineering and biological sciences have lead to
discovery of several noise induced phenomena in systems far away from equilibrium.
Prominent examples are noise induced phase transitions, stochastic resonance, resonant
activation, noise induced stability of unstable states, noise induced unidirectional
transport of particles in the absence of obvious bias( thermal ratchets or theory of
molecular motors), noise induced ordering ranging from separation of different materials
into heterogeneous final state to the formation of  a rich variety of regular patterns,
self organization, etc[1-8]. The list is not exhaustive by any means. In all these
phenomena mentioned above noise plays an active role( active noise paradigm). Noise or
fluctuations arise either because of the coupling of system to an external unknown system
or  from the thermal bath.
The presence of noise can alter the behaviour of system in a fundamental way - a change that
has nothing to do with the sensitive dependence of initial conditions as in the chaotic
systems. In contrast to the general notion that noise is undesirable and destructive, in
many nonequilibrium systems it plays a constructive and stabilizing role in the dynamics.
The most of the phenomena mentioned above exist only in the presence of noise, i.e., they
can not be observed in the absence of noise. Hence noise acts as a generator of order as
opposed to generator of disorder. Even when the magnitude of noise is small, the
probability of the macrostate depends on the details of the global kinetics and can not be
determined by the macrostate alone. In other words the stability criteria which examine only the immediate vicinity of a locally stable state are inadequate to assess the relative stability of states in a nonequilibrium system. The kinetics of unstable intermediate state even as these are rarely populated can have a dramatic effect on the relative stability of states. This leads to the notion of
local versus global stability criteria that will be  discussed in the following sections.

In the present work we will be mainly concerned with the nature of directed motion induced by the
random noise in inhomogeneous systems in the absence of bias. The noise induced active transport in a fluctuating
environment arises from so-called ratchet mechanism[7-24]. Here, nonequilibrium fluctuations
combined with spatial or temporal anisotropy conspire to generate systematic motion even in the
absence of any net bias. It should be noted that in accordance with the second law of
thermodynamics usable work cannot be extracted if only equilibrium fluctuations are present.
In  thermal equilibrium the principle of detailed balance prohibits net particle
current in any system. In contrast, in a
nonequilibrium situation, where the detailed balance is lost, net current flow is possible,
i.e., one can extract energy by rectifying fluctuations or at the expense of overall
increased entropy. The problem of rectification at Brownian scale was posed much earlier
by Smoluchowski[25] and Feynman[26]. A major motivation for these studies(stimulating interaction between
biology and physics) comes from the plausible theoretical
arguments for the motion of molecular kinesin[1,7,8,11]. The kinesin molecule
belongs to a class of 
proteins known as motor molecules. These molecules which include
dyneins and myosin move along 
the structural filaments such as microtubules, microfilaments. The
motor molecules are used for the 
transportation of organnelles(cargo, chemicals) for intra cellular
transport and muscle contraction or to power muscles. 
The energy source for these molecules comes from hydrolysis of ATP.  
In ATP energy is stored in the phosphate bonds, this energy is
released when the bond is hydrolyzed 
and ADP is produced.  The motor proteins use this energy to bring
about unidirectional motion 
along the biopolymers. Here chemical energy is converted into
mechanical energy. The fluctuations 
in the potentials experienced by the motors are believed to arise from
the binding and dissociation 
of ATP, and the anisotropic periodic potential as representing the
electrostatic potential 
along the long structural filament. In these systems it is known that
there is no known gradient 
of chemical concentration or the temperature, to determine the
direction of the movement. 
Moreover, it is also clear that the motion of these motors can be
described as a over damped motion of 
a Brownian particles. At any time the velocity $v$ of the particle is
proportional to the force $F$ on 
the particle. These particles  experience random kicks from the
surrounding medium 
and the average thermal energy of a particle is $k_{B}T$. This energy
$k_{B}T$ is comparable to the other involved 
energy scales in the problem, such as the barrier heights.
Hence Brownian motion plays an essential role in the action of
motors. Thus, protein motors operate in a Brownian regime where
inertia is negligible and thermal fluctuations are important.

With a primary motivation to explain the unidirectional motion of molecular motors the subject of
noise induced transport has gone beyond the biological realm.  In physics, area of thermal ratchets
are being explored to investigate new methods  for controlled devices of high resolution
for particle separation[27]. These devices are expected to be
superior to existing methods such as electrophoretic method for particles of micrometer scale,
like cells, latex spheres, DNA or proteins. For this the current
reversal phenomenon 
is one of the most interesting aspect of the theory of Brownian
ratchets, i.e., magnitude and the 
direction of the Brownian particles are very sensitive to their masses.
New questions regarding the nature of heat engines (reversible and irreversible)
at atomic or molecular scales, their energetics,
and the efficiency of the energy conversion are being studied[28]. In Brownian heat engines one would
also like to understand the possible sources of irreversibility and whether the irreversibility
can be suppressed such that the efficiency approaches that of a Carnot cycle.

The  collection of large number of interacting Brownian particles exhibit
cooperative effects.  Interactions can lead to dynamical phase transitions and instabilities that
characterise the behaviour of motor collections. It turns out that cooperative motors can
generate a directed force even if the system is symmetric[29]. The direction of motion of a symmetric
system is selected by spontaneous symmetry breaking. In symmetric case, the transition is
isomorphous to a paramagnet-ferromagnet transition, in the asymmetric case
to a liquid-vapor transition. In some special cases the coupled systems display amazing features
when exposed to an externally applied force. The most prominent being a zero-bias
negative conductance and anomalous hysteresis[30], which are completely new collective phenomena.
In game theory based on the theory of Brownian ratchets, new area of paradoxical gambling games
have emerged under the subject of Parronodo's paradoxes[31]. Here, two losing (with probability one)
gambling games, when combined or played in a random sequence, can lead to a winning
game with probability one!  It has also been suggested that mobility-induced transport
can be used as a concept of possibilities to think about social and financial systems that are so often
out of equilibrium[32]. Ratchet devices based on quantum processes have been proposed. Using quantum dots these
devices have been investigated experimentally. Net currents due to quantum coherent motion of electrons
have been observed even when the applied
ac voltage is zero on average. Strikingly the direction of the current is reversed as a function of
temperature, exhibiting quantum to classical cross over, the phenomenon is of fundamental
importance in the foundations of quantum mechanics[33-34].

Several qualitatively different physical models for noise induced transport have been proposed, namely,
rocking ratchets, flashing ratchets, diffusion ratchets, correlation ratchets, frictional ratchets, etc.
For this we refer the readers to an excellent recent review by Reimann[8]. Most of these
studies are restricted to non-interacting particles. In this article we confine to the physics in
inhomogeneous medium where Brownian particles experience the friction coefficient and/or temperature to
be non uniform in space. We find out a model with minimal conditions that the inhomogeneous system
assisted by fluctuating forces must satisfy in order to obtain unidirectional current. For this we develop
a self consistent approach based on microscopic treatment[24,35]. In the process we discuss about thermal and
frictional ratchets that are of current interest. In these inhomogeneous systems we show that
directed current can be obtained even in spatially periodic symmetric potentials. The inversion symmetry in
these systems being broken dynamically by the space dependent
frictional coefficient or temperature inhomogeneity. Several other
conditions to obtain noise induced transport can be relaxed in regard to the nature of external noise
and their statistics as compared to non frictional ratchets. We explain the possibility of obtaining
current reversals even in the adiabatic regime of rocked frictional ratchets as a function of system
parameters[36,37]. In a non-adiabatic regime multiple current reversals can be observed[38].
 Following the treatment of stochastic energetics we discuss the efficiency of the energy
transduction and give conditions under which noise can facilitate the energy conversion.

Nature of system inhomogeneity plays important role in deciding the nonequilibrium and
kinetic properties of the system. Most of the systems that one comes across in nature are
inhomogeneous. These inhomogeneities could be structural, configurational, entropic,
temperature non-uniformities, etc. Brownian motion in confined geometries or in porous media
show space dependent friction[39]. Particles diffusing close to surface have a space dependent
friction coefficient[39,40]. It is believed that molecular motor proteins move close along
the periodic structure of microtubules and therefore experience a position dependent mobility[32].
Frictional inhomogeneities are common in  super lattice structures and in semiconductor systems[20]. In Josephson junctions
periodically varying frictional coefficient corresponds to the term representing interference
between  the quasiparticle tunneling and the Cooper pair tunneling[41]. Nonuniformity in temperature
can have important consequences on the particle motion, for instance, the kinetics of growth of
crystalline nuclei in the melt around its critical size. The latent heat generation being, 
in this example, responsible for the creation of nonuniform temperature field across the surface
of the nucleus. One can have inhomogeneous temperature field because of nonuniform 
distribution of phonons and electrons(or of quasiparticles in general) with different
characteristic temperatures in the solid[22]. Temperature inhomogeneities can also  be induced
by external pumping of noise in the system.

We would like to emphasize and show explicitly later that the variations in space dependent
friction does not alter the equilibrium properties of the system, however, affects the dynamics
of the system in a nontrivial way. The relative stability of the competing states is generally
governed by the usual Boltzmann factor. In the presence of nonuniform temperature the
relative stability of two states is sensitive to  detailed
kinetics all along the pathways 
(on the potential surface between the two states under comparison).
We reason in the present work that system 
inhomogeneity may provide a clear and unifying framework to
approach the problem of macroscopic motion under discussion. 
The existing popular models,[8-19] currently in the literature, mostly 
take the nonequilibrium fluctuations to be  non-Gaussian white (colored) noise
together with a ratchetlike periodic potential to aid
unidirectional motion of an overdamped Brownian particle.
The ratchetlike periodic system potentials $V(q)$, 
obviously violate parity $V(q)\neq V(-q)$. 
For such  potentials  one can readily calculate
steady current flow $J(F)$ of a Brownian particle in the
presence of an external field $F$. It turns out that $J(F)$ is
not an odd function of $F$ and in general, $J(F)\neq - J(-F)$ in the regime of nonlinear response. 
In other words, reversal of the external force may not lead to a reversed
current of the same magnitude in sharp contrast to the case of a
nonratchetlike (symmetric) periodic potential system where
$J(F)$= $-J(-F)$ follows. From this general observation, in a
ratchetlike potential, it can be easily concluded that on
application of a zero time averaged periodic field, say
$F$=$Asin\omega t$, one can obtain net unidirectional current.
Thus ratchet may also be described as a nonlinear rectifier.
The currents exhibit maxima both as a function of the forcing
amplitude as well as temperature. When the forcing amplitude is small
the motion is determined by the thermal activation rates ( Kramer's
time scales) or overcoming the barriers to the left and right. Jump
rates being asymmetric, we obtain a finite average drift velocity for
nonzero forcing term. The average velocity is quite small at low
temperatures due to the Arrehenius prefactor. At intermediate
temperature, the jumps are more frequent and at high temperatures,
jumps in both direction are of equal likelihood and the magnitude of the
net current falls. For this reason we observe a maxima in current at
finite temperature. 
 This is the basic physics behind 
some of the physical models (known as rocked ratchet) used to obtain
current rectification in a periodic potential system.
There are models, however, that do not
use oscillating external fields. Instead, colored noise
of zero average strength----dichotomous,
Ornstein-Uhlenbeck, Kangaroo processes,...[11-13], is used to drive the
Brownian particle to obtain macroscopic motion in a ratchetlike
potential system. We would like to emphasize that in rocked ratchet it 
is possible to obtain unidirectional current even in the absence of
thermal noise, i.e., in the deterministic limit when the potential is
asymmetric and the amplitude of the rocking force can be such that
tilt in one direction can destroy all energy barriers allowing the
particle to slide down ( running state). However, they still have barriers when
the potential is tilted by the same amount in the opposite direction and hence preventing 
the particle motion in the opposite direction. In this deterministic
limit, motion exhibit intriguing structures as the function of forcing
amplitude, such as current quantization and phase locking behaviour[38]. Inclusion
of inertial effect allows the possibility of having both regular and
chaotic dynamics. This deterministically induced chaos can mimic the
role of noise. The system exhibits multiple current reversals with
respect to the forcing amplitude. However, all these effects are not
robust in the presence of noise.

There are further interesting models where the
potential barriers themselves are allowed to fluctuate (flashing ratchet), for
instance, with finite time correlations between two states under
the influence of a noise source[8,11]. An example
being an overdamped Brownian particle subjected to a ratchetlike
periodic potential, where the asymmetric saw-tooth potential (in
which two edges of the teeth make different angles with the vertical) is 
switched on to its full strength for time $\tau_{on}$ during which
the Brownian particle slides down the potential slope to the bottom of
the potential minima (fixed point). At the end of $\tau_{on}$, the system is put in
the other ($off$) state during which the potential is set equal
to a constant (say = 0, simplest case)
for an interval $\tau_{off}$ and the particle executes
free diffusive motion. At the end of $\tau_{off}$ the system is put back
in the $on$ state for interval $\tau_{on}$. This process of flipping of
states is repeated ad-infinitum. If $\tau_{off}$ is adjusted in such
a way that by the end of $\tau_{off}$ the diffusive motion just takes 
the particle out of the (now nonexistent) potential minima in the 
steeper slope direction (smaller distance, say in right direction) of
the saw-tooth potential 
but fails to do so in the gentler slope direction (larger distance),
the immediate next $on$ interval will take the particle to the adjacent 
minimum in the steeper slope side of the saw-tooth potential.
Repetition of such sequential flipping of states 
for a large number of times lead to a net unidirectional
macroscopic current (in the right direction) of the Brownian particle. It should be noted that
a symmetrical nonratchetlike potential(fluctuating between two states) would, instead, have yielded symmetrical
excursions of the Brownian particle and, hence, no net unidirectional motion.
In this mechanism to obtain net current of the 
Brownian particle the system is supplied
with the required energy externally to flip the system between the two states 
keeping the interval $\tau_{on}$ and $\tau_{off}$ fixed ( thus the
system is in a nonequilibrium state). There is a
lot of freedom to play around with the parameters $\tau_{on}$, $\tau_{off}$
, the saw tooth potential, and the thermal 
noise strength. And a judicious tuning of these parameters could 
even result in the reversal of the current. The flipping process could,
however, be effected also by a finite time-correlated 
fluctuating dichotomous force.
It should be noted that the former flipping 
process of definite $\tau_{on}$ and 
$\tau_{off}$ have been practically exploited in the particle separation
techniques, whereas the latter fluctuating flipping time process
has some appeal to natural processes.
The diversity of the models just
does not end here. Unlike rocking ratchets ( deterministic limit with
high amplitude of rocking), thermal fluctuations are essential to get net 
unidirectional motion in all models of flashing
ratchets. Interestingly in all these flashing ratchet models particle
need not climb 
the barrier, it only has to slide down the potential slope. There have
been attempts too to obtain 
macroscopic current with Gaussian white noise under nonratchetlike
symmetric periodic potential field as well but subjected to
temporally asymmetric periodic external fields [16-18]. We do not attempt
here, however, to give a review of the models considered. We
present, in the following, a framework to obtain macroscopic
motion in an inhomogeneous system with space dependent friction
coefficient and nonuniform temperature fields restricting
ourselves to periodic potential systems.
\par In all the models just mentioned [8] (The list is not
exhaustive.) the system was
taken to be homogeneous as far as the question of diffusivity
was concerned. However, in some of the works, earlier to the
ones alluded to so far, the nonuniformity of the diffusion constant
of the system was considered to yield macroscopic transport [20-21 ].
The diffusion coefficient could be space dependent or
state dependent and so the system may dissipate energy during
its time evolution differently at different space points.
Unlike homogeneous systems,
however, the physics of inhomogeneous systems has not been free
from controversies[21,44], such as, whether the equation
\begin{subequations}
\begin{eqalignno}
\frac{\partial P}{\partial t}=\frac{{{\partial}^2}}{\partial q^{2}}
{D(q)P},
\end{eqalignno}
or \\
\begin{equation}
\frac{\partial P}{\partial t}=\frac{\partial}{\partial q}{D(q)}
\frac{\partial P}{\partial q},
\end{equation}
\end{subequations}
\\
\noindent should be the correct form of diffusion equation. Nevertheless,
such controversies apart, B$\ddot u$ttiker [20] and also van
Kampen [21] have shown that one can expect macroscopic transport
of a Brownian particle in a periodic potential field when the
diffusion coefficient is also periodic with the same periodicity
but shifted by a phase difference other than 0 and $\pi$ with
respect to the periodic potential field. It should be noted that
the potential field is not required to be ratchetlike. The
system is rendered nonequilibrium by diffusion coefficient
inhomogeneity in the system and the "stationary state" of the
system is no longer governed by the usual Boltzmann factor.
In the theory, one needs to go beyond the
phenomenological description.

In reference [35], a microscopic treatment is given for the
derivation of the macroscopic equations of motion in an
inhomogeneous medium (space dependent friction coefficient and
spatially nonuniform temperature) starting from a microscopic
Hamiltonian of the system in contact with (phonon) heat bath(s).
Moreover, a proper overdamped limit of the Langevin equation of
motion in such an inhomogeneous medium is derived. 
A correct form of the corresponding Fokker-Planck equation is 
obtained and it is
explicitly shown that neither of the two forms of the diffusion
equation mentioned above [eq.(1)] is correct.
From this macroscopic
equation of motion one obtains an expression for the average
current which depends on the details of the potential field and
the inhomogeneities of the system.

As mentioned earlier, the nonuniformity of diffusion
coefficient can arise either because of the space dependence of
the friction coefficient $\eta(q)$ or that of the temperature
$T(q)$ or because of both[40.44].
We, however, discuss possibilities of
macroscopic current flow as a result of various kind of
inhomogeneities in a symmetric periodic potential
system. The first case we consider is when $\eta(q)$ and $T(q)$
are space dependent. 
By taking $\eta(q)$ and $T(q)$
periodic one gets a tilt in the potential field throughout the
sample as discussed phenomenologically in ref.[20],
resulting in a macroscopic current because of thermal
fluctuations. In this case the temperature inhomogeneity is
crucial and one can obtain current even when the friction
coefficient becomes uniform. Friction coefficient
inhomogeneity alone, however, does not generate macroscopic current. In
the second case, we consider a thermal particle in a system with
space dependent friction coefficient but subjected to 
external white noise fluctuations, and in the third 
case a thermal particle subjected to external space dependent
white noise is considered. We then discuss the case of
a Brownian particle coupled to two thermal baths.
It should be noted, however, that in
all the cases that we have considered  do not require
the potential to be ratchetlike nor do we require the
fluctuating forces to be correlated in time to obtain
macroscopic current. Finally we discuss the
noise induced currents, their efficiency and the phenomenon of
multiple current reversals in 
rocked frictional ratchets.

\par In section 2 we provide a derivation of the
macroscopic equation of motion in an inhomogeneous system from a 
microscopic Hamiltonian of a Brownian particle interacting with a
(phonon ) heat bath. We, then, obtain proper Smoluchowski equation
from the derived Langevin equation of motion following the
prescription of Sancho et al [45]. We use, in Sec.3, this
overdamped equation of motion in an inhomogeneous system with space
dependent friction coefficient and nonuniform temperature field to 
obtain nonzero macroscopic current. In the same section we elaborate
three other possible cases of inhomogeneous systems where macroscopic
current could be possible.
The section 4 is devoted to AC driven frictional ratchets and 5 for 
discussions.

\section{Equation of motion in inhomogeneous systems}

We consider an inhomogeneous system where the inhomogeneity
could arise either because of the space dependence of friction
coefficient, or the nonuniformity of the temperature field or
because of the combined effect of both. The effect of the
nonuniformity of temperature or temperature gradient, however,
cannot be incorporated as a potential term in the Hamiltonian
formalism in sharp contrast to, for instance, the amenability of
incorporation of electric field gradient in the Hamiltonian of a
charged particle. We, therefore, incorporate the effect of
temperature inhomogeneity at the end directly into the equation 
of motion obtained
from the microscopic Hamiltonian suited to take care of the
space dependence of the friction coefficient.

\subsection{Equation of motion in a space dependent friction field}

We consider a (subsystem) Brownian particle, of mass $M$,
described by a coordinate $Q$ and momentum $P$ moving in a potential
field $V(Q)$ of the system and being in contact with a
thermal(phonon) bath. The bath oscillators are described by
coordinates $q_{\alpha}$, momenta $p_{\alpha}$ and mass $m_{\alpha}$ 
with characteristic frequencies $\omega_{\alpha}$. We consider
the total Hamiltonian[35]
\begin{equation}
H \: = \: \frac{P^2}{2M}+V(Q)+\sum_{\alpha}\left[\frac{p_{\alpha}^2}{2m_{\alpha}}
\: + \: \frac{m_{\alpha}\omega_{\alpha}^2}{2}
\left(q_{\alpha} \: - \: \lambda_{\alpha}\frac{A(Q)}
{m_{\alpha}\omega_{\alpha}^2}\right)^{2} \right],
\end{equation}
\noindent The interaction of the subsystem with the thermal bath
is through the linear (in $q$) coupling term
$\lambda_{\alpha}q_{\alpha}A(Q)$. From (2) one obtains the
following equations of motion.
\begin{subequations}
\begin{eqalignno}
{\dot Q} \: = \: \frac{P}{M} ,
\end{eqalignno}
\begin{eqalignno}
{\dot P} \: = \: - V^{\prime}(Q)+\sum_{\alpha}\lambda_{\alpha}
A^{\prime}(Q)\left[q_{\alpha} \: - \: \lambda_{\alpha}
\frac{A(Q)}{m_{\alpha}{\omega_{\alpha}^2}}\right],  
\end{eqalignno}
\begin{eqalignno}
{\dot q}=\frac{p_{\alpha}}{m_{\alpha}},
\end{eqalignno}
\begin{equation}
{\dot p_{\alpha}} \: = \: - m_{\alpha}{\omega_{\alpha}^2}q_{\alpha}
\: + \: \lambda_{\alpha}{A(Q)},
\end{equation}
\end{subequations}
\noindent where $A^{\prime}(Q)$ is the derivative of $A(Q)$ with respect
to $Q$. After solving (3c) and (3d) for $q_\alpha$ by using the
method of Laplace transform and substituting its value in (3b),
we obtain
\begin{subequations}
\begin{eqalignno}
{\dot Q} \: = \: \frac{P}{M},
\end{eqalignno}
\begin{eqnarray}
{\dot P(t)} \: = \: - \: V^{\prime}(Q) \: - \:
\sum_{\alpha}\frac{\lambda_{\alpha}^{2}
A^{\prime}(Q)}{m_{\alpha}\ {\omega_{\alpha}^2}}
\int_{0}^{t}dx \: {\cos\omega_{\alpha}(t - t^{\prime}) A^{\prime}(Q)
{P(t^{\prime})\over M}}  \nonumber\\
+A^{\prime}(Q) \sum_{\alpha} \lambda_{\alpha}
\left[ x_{\alpha}(0)
\cos(\omega_{\alpha}t)
\: + \: {{\dot x}_{\alpha}(0) \over \omega_{\alpha}}
\sin(\omega_{\alpha}t) \right]
+A^{\prime}(Q) \sum_{\alpha} \frac{A(Q_0) \lambda^{2}_{\alpha}}
{m_\alpha \omega^{2}_\alpha} \cos(\omega_\alpha t). \yesnumber
\end{eqnarray}
\end{subequations}
\noindent Here $Q_0$ is the initial value of the particle
co-ordinate $Q$ and $x_\alpha (0)$ and $\dot x_\alpha (0)$ are
the initial co-ordinates and velocities, respectively , of the
bath variables.
The second term in the right hand side of equation (4b)
depends on the momenta at all times
previous to $t$. At this stage Markovian limit is imposed so that
\begin{equation}
g(t - t^{\prime}) \: = \: \sum_{\alpha}\frac{{\lambda_{\alpha}^{2}}} 
{m_{\alpha}\omega_{\alpha}^2}\;
\cos{\omega_{\alpha}(t - t^{\prime})} = 2\eta \delta(t-t^{\prime}).
\end{equation}
\noindent The equation (5) follows readily from the well known
Ohmic spectral density distribution for the bath oscillators, i.e., 
\begin{equation}
\rho(\omega) \: = \: \frac{\pi}{2} 
\sum_{\alpha} \frac{\lambda^{2}_{\alpha}}{m\omega_{\alpha}}
{\delta(\omega - \omega_{\alpha})}= 
\eta\omega{e^{-{\frac{\omega}{\omega_{c}}}}}. 
\end{equation}
\noindent where $\omega_c$ is an upper cutoff frequency set by
the oscillator spectrum of the thermal bath.
The Markovian approximation (5) has the effect of 
neglecting the transient terms involving the initial coordinate $Q_0$, in
the equation of motion, for long time behaviour [46]. 
In other words, the equation should well describe
the motion of the Brownian particle in time scales $t >$
$\omega_{c}^{-1}$.
The equation of motion thus assumes the form
\begin{subequations}
\begin{eqalignno}
{\dot Q} \: = \: \frac{P}{M},
\end{eqalignno}
\begin{equation}
{\dot P} \: = \: - V^{\prime}(Q)  -  \frac{\eta}{M}
[A^{\prime}(Q)]^2{P} + A^{\prime}(Q)\ f(t), 
\end{equation}
\end{subequations}
\noindent where 
\begin{equation}
f(t) \: = \: \sum_{\alpha}\lambda_{\alpha}\left[q_{\alpha}(0)
\cos(\omega_{\alpha}t)
\: + \: \frac{\dot q_{\alpha}(0)}{\omega_{\alpha}} 
\sin(\omega_{\alpha}t) \right].
\end{equation}
\noindent The force $f(t)$ is fluctuating in character because of
the associated uncertainties in the initial conditions
$q_{\alpha}(0)$ and ${\dot q_{\alpha}(0)}$ of the bath variables.
However, as the thermal bath is characterised by its temperature
$T$, the equilibrium distribution $P_{eq}(q_{\alpha}(0),\dot q_{\alpha}(0))$
of bath variables is given by
the Boltzmannian form 
$$ \hskip 1.4 in P(q_{\alpha}(0),\dot q_{\alpha}(0))= \frac{1}{Z}
\prod_{\alpha} e^{- \frac{1}{2k_{B}T}
\left( m_{\alpha} \dot q_{\alpha}^{2}(0)
+m_{\alpha} \omega_{\alpha}{^2}  q_{\alpha}{^2}(0) \right)},
\hskip 1.38 in (8a) $$
\noindent where $Z$ is the partition function. Using equation (8a)
and (6) one can easily compute the statistical properties of the
fluctuating force $f(t)$. It is Gaussian with
\begin{subequations}
\begin{eqalignno}
\langle f(t) \rangle = 0,
\end{eqalignno}
\noindent and \\
\begin{eqalignno}
\langle f(t)f(t^{\prime}) \rangle = k_{B}T\ g(t - t^{\prime})
 = 2k_{B}T\eta \delta(t - t^{\prime}).
\end{eqalignno}
\end{subequations}
\noindent It should be noted that the effect of the interaction
term $\lambda_{\alpha}q_{\alpha}A(Q)$ in the Hamiltonian (2) is
to introduce a friction term and a fluctuating term $f(t)$ in the
equation of motion (7b). Moreover, $A^{\prime}(Q)$=constant
corresponds to a uniform friction coefficient. We redefine,
${[A^{\prime}(Q)]^2}\eta = \eta(Q)$ and $\frac{f(t)}
{\sqrt T} \rightarrow f(t)$ , and put $M=1$, in (7) to obtain,
\begin{subequations}
\begin{eqalignno}
{\dot Q} = P,
\end{eqalignno}
\begin{eqalignno}
{\dot P} = - V^{\prime}(Q) - \eta(Q)P + \sqrt{k_{B}{T}{\eta(Q)}} f(t),
\end{eqalignno}
\end{subequations}
\noindent with
\begin{subequations}
\begin{eqalignno}
\langle f(t) \rangle = 0,
\end{eqalignno}
\noindent and \\
\begin{eqalignno}
\langle {f(t)f(t^{\prime})} \rangle = 2 \delta(t - t^{\prime}).
\end{eqalignno}
\end{subequations}
\noindent From eqs.(9) it follows that the derived Langevin
equation of motion (10b) of a Brownian particle, in a system with
space dependent friction $\eta(Q)$ but at constant uniform
temperature $T$, is internally consistent and obeys fluctuation-dissipation
theorem. We now proceed to incorporate the effect of
space dependence of temperature, in a thermally nonuniform
system, into the Langevin equation of motion by assuming that
the Brownian particle comes in contact with a continuous
sequence of independent temperature baths as its coordinate $q$
changes in time. (For notational simplicity, we replace the
coordinate $Q$ and momenta $P$ by the corresponding lower case
letters $q$ and $p$, respectively, reserving $P$ for probability
distribution.)

\subsection{Equation of motion in a space dependent friction and
temperature field}

We consider each space point $q$ of the system to be in
equilibrium with a thermal bath characterised by temperature
$T(q)$. Also, it should be noted that one could take $\eta(q)$ to
be constant piecewise along $q$ , and in each piece of these $q$
segments eq.(10b) would correspond to an equation of motion with
the constant friction
coefficient but with the same statistical character 
of $f(t)$ (11a-11b) in all $q$ intervals. Let us discretise
the system, for the sake of argument, into segments $\Delta{q}$
around $q$ and represent them by indices $i$. Let us further assume
that each segment is connected to an independent thermal bath at
temperature $T_i$ with corresponding random forces $f_{i}(t)$ so
that the equation of motion (10b), in the segment $i$, will have
the last term $\sqrt{k_{B}T_{i}{\eta(Q)}}f_{i}{(t)}$. As the two
different segments $i$ and $j$ are each coupled to an independent
temperature bath we have $\langle
f_{i}(t)f_{j}(t^{\prime})\rangle = 2 \delta_{ij}\delta(t - t^{\prime})$.
Because $f(t)$ is $\delta$ correlated in time, as the particle
evolves dynamically the fluctuation force $f_{i}(t)$ experienced
by the Brownian particle while in the space segment $i$ at time $t$
will have no memory about the fluctuating force experienced by
it at some previous time $t^{\prime}$ while in the space segment
$j \neq i$. The space-dependent index $i$ in $f_{i}(t)$,
therefore, can be ignored and the equation of motion becomes
local in time as well as in space. Therefore, in the continuum
limit, the stochastic equations of motion of the Brownian
particle, in an inhomogeneous medium with space dependent
friction and nonuniform temperature, acquire the simple forms
\begin{subequations}
\begin{eqalignno}
{\dot q} = p,
\end{eqalignno}
\begin{eqalignno}
{\dot p} = - V^{\prime}(q) - \eta(q)p + \sqrt{k_{B}T(q)\eta(q)} f(t),
\end{eqalignno}
\noindent with
\begin{eqalignno}
\langle f(t)f(t^{\prime}) \rangle = 2 \delta(t - t^{\prime}).
\end{eqalignno}
\end{subequations}

\subsection{The Smoluchowski Equation}

From eq.(12b) one can readily write down the Fokker-Planck
equation or the Kramer's equation for the full probability
distribution $P(q,v,t)$. However, in most of the practical
situations the marginal probability distribution $P(q,t)$ for the
variable $q$ alone suffices to describe the motion of the
Brownian particle. This probability distribution $P(q,t)$ can be
obtained in the overdamped limit of the Langevin equation (12b)
which is valid on time scales larger than the inverse friction $\eta^{-1}$.
In other words in the overdamped case the fast
variable, velocity $v$, is eliminated from the equation of motion. 
In the case of
homogeneous systems one simply puts ${\dot p} = 0$ in eq.(12b) to obtain the 
overdamped Langevin equation. However, in case of inhomogeneous systems, the
above method of adiabatic elimination of fast variables 
does not work, and leads to unphysical equilibrium distribution. 
The proper prescription for the elimination of
fast variables has been given in Ref.[45] 
for systems with space dependent friction. 
The method retains all terms upto order $\eta^{-1}$ 
and the resulting overdamped Langevin equation yields 
physically valid equilibrium distribution. We,
therefore, apply the same prescription to obtain the overdamped 
Langevin equation
of motion in an inhomogeneous system with space dependent friction $\eta(q)$ 
 and nonuniform temperature field $T(q)$. We obtain,
\begin{equation}
{\dot q} = - \frac{V^{\prime}(q)}{ \eta(q)} -
\frac{k_{B}}{2 {[\eta(q)]}^2}
\left[T(q) \eta^{\prime}(q) + \eta(q)T^{\prime}(q) \right] + 
\sqrt{ \frac{k_{B}T(q)}{\eta(q)}}f(t),
\end{equation}
\noindent with \\
\begin{equation}
\langle f(t)f(t^{\prime}) \rangle \: = \: 2 \delta(t - t^{\prime}).
\end{equation}
\noindent Using van Kampen Lemma[47] and the Novikov's
theorem[48] we obtain the
corresponding Fokker-Planck equation as
\begin{equation}
{\partial P(q,t) \over \partial t} \: = \: {\partial \over \partial q}
{1 \over \eta(q)}\left[{\partial \over \partial q}k_{B}T(q)P(q,t) \: + \:
V^{\prime}(q) P(q,t)\right].
\end{equation}
\noindent Eq.(15) is the Smoluchowski equation for an 
overdamped Brownian particle
moving in an inhomogeneous system with space dependent friction and 
nonuniform temperature. It should be noted that eq.(15) 
gives the correct form of
diffusion equation instead of either of the two forms mentioned in eqs.(1).
It is clear that the temperature and the friction coefficients influence the 
particle motion in a qualitatively different fashion and they cannot be 
plugged
together to get effective diffusion coefficient to satisfy either of the forms
of eq.(1). In the next section we discuss how the
system inhomogeneity can 
help maintain a macroscopic unidirectional current.

\section{Unidirectional currents in inhomogeneous systems (frictional ratchet)}

We consider inhomogeneous systems where the inhomogeneity could be an
internal property of the system or it could be imposed externally. As mentioned 
earlier we consider four cases where macroscopic motion can be obtained.

\subsection{Rectification in an inhomogeneous 
system with space dependent 
friction and nonuniform temperature}
\par When the system is bounded at $q \rightarrow \pm \infty$, i.e.,
$V \rightarrow \infty $ as $q \rightarrow \pm \infty$, the system
attains steady (stationary) state with zero probability current.
In such a situation, we can calculate the steady state
probability distribution $P_{s}(q)$, from the Smoluchowski
equation (15), by setting the probability current
\begin{equation}
{1 \over \eta(q)}\left[V^{\prime}(q)P(q,t) + {\partial \over \partial q} 
k_{B}T(q)P(q,t)\right]
\end{equation}
\noindent equal to zero, as 
\begin{equation}
P_{s}(q) = {\it N} e^{-\psi(q)},
\end{equation}
\noindent where\\
\begin{equation}
\psi(q) \: = \: \int^{q} \left(V^{\prime}(x) + 
kT^{\prime}(x) \over k_{B}T(x)\right) dx,
\end{equation}\\
\noindent and {\it N} is a normalization constant. 
It is very clear from the expression, eq.(18), for $\psi (q)$
(effective or generalized potential)
that the peaks of $P_{s}(q)$ are determined not by the minima
of $V(x)$ alone but are determined as a combined effect with
$T(x)$. $P_{s}(q)$ may even peak at positions which would be
quite less likely to be populated in the stationary situations
for uniform temperature, $T(x)$ = $T$ condition.
In this respect, nonequilibrium situations appear strange.
It is quite common in biological systems where, for example, otherwise
less likely ion channels are, in some situations, found to be
more active for ionic transport. Recently such nonequilibrium
behaviour in biological systems have been theoretically
attributed to the effect of nonequilibrium fluctuations and the
process has been termed as kinetic focusing[49]. Moreover, it
should be noted that the relative stability of two states of a
system with nonuniform temperature field is not determined by
the local function $V(q)$ but by the entire pathway through a
continuous sequence of intervening states between the two states
under comparison. The
temperature variation may modify the kinetics of these
intervening states drastically and hence their contribution
towards the relative stability will be substantial even when they are
sparsely populated. For example, application of a localized heating at
a point on the reaction coordinate lying between the lower energy
minimum and potential energy barrier maximum can raise the relative
population of the higher-lying energy minimum over that of a lower
minimum given by usual
Boltzmann factor. 
It should further be noted 
that $\psi(q)$ is not determined by $\eta(q)$ as it should be.
Moreover, the functional form of $\psi(q)$ is similar to
$\int^{q}{v(x) \over D(x)}dx$, of course, in this case $V^{\prime}(q)$ 
has been augmented by a compensating force k$T^{\prime}(q)$.
$v(q) = \eta^{-1}(q)[V^{\prime}(q)+kT^{\prime}(q)]$ is the drift 
velocity and D(q) = $\eta^{-1}(q)k_{B}T(q)$ is the effective diffusion
coefficient.
\par
\begin{figure}
  \centerline{\epsfysize=14cm \epsfbox{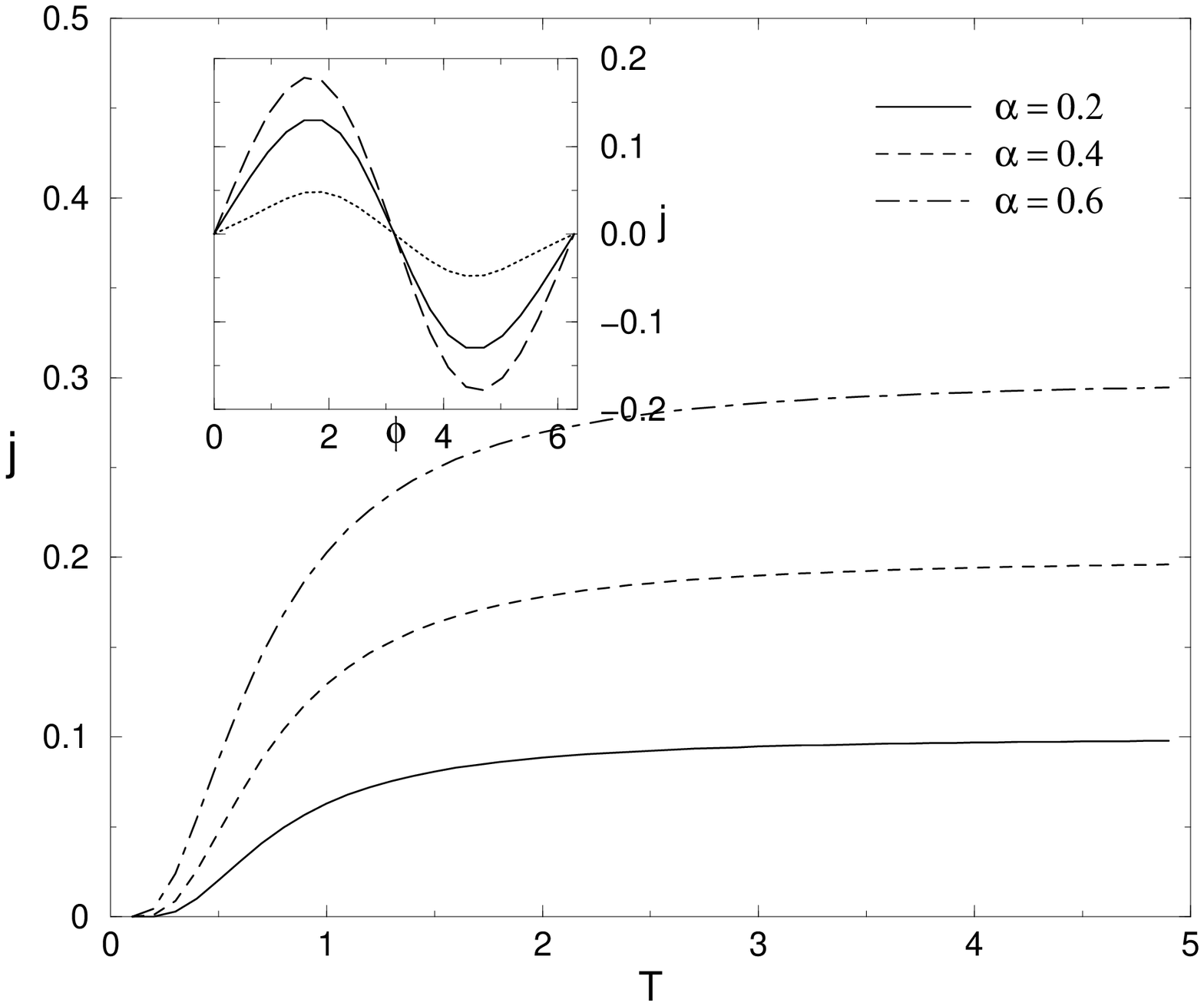}}
  \vspace{.7in}
  \caption{Current is plotted as function of temperature $k_{B}T_{0}$ for
   $\phi = \frac{\pi}{2}$ and $\alpha=0.5$. The inset shows current
  as function of $\phi$ for $k_{B}T_{0}=2.0, k_{B}T_{0}=1.0, k_{B}T_{0}=0.5$ 
    from top to bottom and $\alpha=0.4$. }
  \label{Butt}
\end{figure}

 So far we have not assigned any functional form to $V(q), T(q)$
and $\eta(q)$.
In ref.[20] it is shown that at least in one case the system can
generate nonzero probability current, namely, when both $V(q)$
and $D(q)$ are periodic with same periodicity but having a 
phase difference other than 0 and $\pi$. 
In our present problem if we assume $V(q)$, $T(q)$ and $\eta(q)$
to be
periodic functions with periodicity, say, $2\pi$ then the probability current is
given by [20]
\begin{equation}
J \: = \:  {{1 - e^{-\delta}} \over {\int_{0}^{2\pi}dy e^{-\psi(y)}\:
\int_{y}^{y+2\pi}dx {e^{\psi(x)} \over D(x)}}} ,
\end{equation}
\noindent where $\delta = \psi(q) - \psi(q+2\pi)$ determines the
effective slope of a generalized potential $\psi(q)$ and hence
$\delta$ being $+$ or $-$ve determines the direction of current. It is
obvious from the expression for $\delta$ that the phase difference
$\phi$, between $V(q)$ and $T(q)$, alone determines the direction of
current. The net unidirectional current remains nonzero (finite)
except when $\phi$ is an integral multiple of $\pi$ ($\phi = n \pi$
corresponds to zero effective slope of the generalized potential). It
should further be noted that the amplitude of variation of $\eta(q)$
does not determine the direction of current but does affect the
magnitude of current. A periodic variation of $\eta(q)$ and $V(q)$ but
uniform $T(q)$ will yield no unidirectional current.  For $V(q) =
V_{0} (1 - cos(q))$ and $T(q) = T_{0}(1 - {\alpha \cos(q-\phi)})$,
with $0 < \alpha < 1 $ (for positive temperature) $\delta$ turns out
to be $\frac{2\pi V_{0} \sin\phi}{k_{B}T} \left[ \frac{1}{\sqrt{1-
      \alpha^2}}-1 \right]$ which is definitely nonzero for $\phi \neq
n\pi$, $n = 0,\pm 1 , \pm2$,.......  Thus $\phi$ alone determines the
direction of nonzero current $J$. In this case the periodic variation
of temperature plays the crucial role and may yield current even when
$\eta(q) = \eta_{0}$ = constant. The dependence of current on $\phi$
alone can readily be observed from the fact that the effective
potential $ \psi(q)$ shows a tilt(or finite slope) in positive or in
negative direction depending on the magnitude of $\phi$.  The
transport arises because particles starting from the potential minima
can climb the hot slope more easily than they can climb the cold
slope, created by phase difference between the potential and
temperature profile. Thus transition rates from one value to another
in the right direction is different from that in the left
direction($\phi$ breaks the symmetry between left and right transition
rates). We consider another simple case for which $V(q)= V_{0}(1-
\cos(q))$, $T(q)=T_{0}/(1-\alpha \cos(q-\phi))$, with $0 \leq \alpha <
1$ and space independent friction with magnitude $\eta_{0}$. For this
simple case effective potential $\psi(q) = V_{0}
(1-\cos(q)-(\alpha/4)[\cos(\phi)-\cos(2q-\phi)]-(\alpha/2)\sin(\phi)q)/T_{0}$,
where the last term clearly shows tilt responsible for the current.
The magnitude of current is given by the eqn. (19) with $\delta=
V_{0}\alpha \sin(\phi) /(2T_{0})$. Current is zero for $\alpha = 0$
and for $\phi = n\pi$. Direction of current depends on $\phi$. In fig.
\ref{Butt} we have plotted dimensionless current $j$ versus $T_{0}$
the average value of the temperature field over the spatial period for
various values of $\alpha$. Here temperature is in a dimensionless
form and is scaled with respect to barrier heights. Other parameters
are given in the figure caption. Interestingly the current increases
with $T_{0}$ starting from $0$ at $T_{0}=0$ and saturates to a
constant value at high $T_{0}$ limit. This saturation in the high
temperature limit is specific to this model. In the inset we have
shown the variation of current $j$ with $\phi$ for various values of
$T_{0}$. It may be noted again that the direction of current is
determined solely by $\phi$.  For further analalysis and graphical
presentation of the effective potential we refer to [20].  We now
consider cases, where $\eta(q)$ plays a decisive role.

\subsection{Rectification in an inhomogeneous system with space dependent
friction in the presence of an external parametric white noise}

Unlike the case considered in subsection 3.1, where the
overdamped Brownian particle experiences a fixed (in time) local
(nonuniform) temperature profile $T(q)$ during its sojourn
$q(t)$ for all $t$,
we consider, in this subsection, a system with uniform temperature
$T(q)=T$ but a spatially varying $\eta(q)$.
The Langevin equation of motion is given by
\begin{equation}
{\dot p} = -V^{\prime}(q) - \eta(q) p + \sqrt{k_{B}T \eta(q)} f(t)
\end{equation}
\noindent and the corresponding overdamped equation is
\begin{equation}
{\dot q} = - \frac{V^{\prime}}{\eta(q)} - \frac{k_{B}T\eta^{\prime}(q)}
{2[\eta(q)]^2} + \sqrt{\frac{k_{B}T}{\eta(q)}} f(t),
\end{equation}
\noindent with
$$\langle f(t) \rangle = 0,$$ and
$$\langle f(t)f(t^{\prime}) \rangle = 2 \delta(t - t^{\prime}).$$
We, now, subject the system to an external parametric additive
white noise fluctuating force $\xi(t)$, so that the equation of
motion becomes
\begin{equation}
{\dot q} = - \frac{V^{\prime}}{\eta(q)} - \frac{k_{B}T\eta^{\prime}(q)}
{2[\eta(q)]^2} + \sqrt{\frac{k_{B}T}{\eta(q)}} f(t) + \xi(t),
\end{equation}
\noindent with
$$\langle \xi(t) \rangle = 0,$$
$$\langle \xi(t) \xi(t^{\prime}) \rangle = 2 \Gamma 
\delta(t - t^{\prime}), $$
\noindent where $\Gamma$ is the strength of the external white noise $\xi(t)$.
We can immediately write down the corresponding 
Fokker-Planck (Smoluchowski) equation
\begin{equation}
\frac{\partial P}{\partial t} \: = \:  \frac{\partial}{\partial q}
\left[\   \left\{\frac{V^{\prime}(q)}{\eta(q)} \right\} P 
 + \left\{\frac{k_{B}T}{\eta(q)} + \Gamma \right\}
\frac{\partial P}{\partial q} \right].
\end{equation}
\noindent For periodic functions $V(q)$ and $\eta(q)$, with
periodicity $2\pi$, one obtains unidirectional current following
earlier procedure using equation (23). The resulting expression
for current $J$ takes the same functional form as given in
eq.(19) where $\psi(q)$ is now given by 
\begin{equation}
\psi(q) = \int^{q} dx \frac{ V^{\prime}(x)}{k_{B}T+\Gamma \eta(x)}
\end{equation}
\noindent and the effective diffusion coefficient \\
\begin{equation}
D(q) = (k_{B}T + \Gamma \eta(q))/\eta(q),
\end{equation} 
\noindent with 
$$\delta = \psi(q) - \psi(q+2\pi). $$\\
For V(q) = $V_{0}(1-\cos(q))$ and $\eta(q) = \eta_{0} (1 -
\alpha \cos(q - \phi))$,
$\delta$ turns out to be equal to
$ \frac{2\pi V_{0} \sin\phi}{ \alpha \eta_{0}}
\left[ \frac{k_{B}T+\eta_{0}}{\sqrt{(k_{B}T+\eta_0)^{2}-
(\eta_{0}\alpha)^2}}-1 \right].$
As earlier the direction of current is determined by the phase
difference $\phi$ between the periodic functions $V(q)$ and $\eta(q)$.
\par It is important to notice that there is no way 
one could obtain macroscopic
current in the absence of the external white noise $\xi(t)$. 
This case , however, is similar in
essence to the previous case of nonuniform temperature. 
In the present situation the overdamped Brownian particle is
subjected to an external parametric random noise. The noise
being externally imposed, the system always absorbs energy
(without the presence of corresponding loss factor) [50].
Also, the overdamped
particle moves slowly wherever the friction coefficient $\eta(q)$ is 
large and the possibility of absorption of energy from the external white
noise at those elements $q$ of the system, therefore, is
correspondingly large. Thus, the effective temperature $T(q)$ of
the system is given by $k_{B}T+\Gamma \eta(q)$,
which modulates as $\eta(q)$ varies and
hence the macroscopic current results as in the case 3.1.

\subsection{Rectification in a 
homogeneous system but subjected to an external
parametric space dependent white noise}

The overdamped Langevin equation is ,
\begin{equation}
{\dot q} = -\frac{V^{\prime}(q)}{\eta} + \sqrt{\frac{k_{B}T}{\eta}} f(t),
\end{equation}
\noindent with $\langle f(t) \rangle$ = 0 and 
$\langle f(t)f(t^{\prime}) \rangle$
 = 2  $\delta(t - t^{\prime})$.
\noindent Eq.(26) obeys fluctuation-dissipation theorem and
hence in the absence of any external bias potential there can be
no net current irrespective of the form of the periodic
potential $V(q)$. We now subject the system to an external
multiplicative Gaussian white noise fluctuation. The
corresponding overdamped Langevin equation is given by
\begin{subequations}
\begin{eqalignno}
{\dot q} = -\frac{V^{\prime}(q)}{\eta} + \sqrt{\frac{k_{B}T}{\eta}} f(t)
+g(q) \xi(t),
\end{eqalignno}
\end{subequations}
\noindent where $g(q)$ is an arbitrary function of $q$ and
$\xi(t)$ is a white noise with 
$$\langle \xi(t) \rangle = 0,$$ and
$$\hskip 1.9 in \langle \xi(t) \xi(t^{\prime}) \rangle = 2\Gamma
\delta(t-t^{\prime}).  $$ \\
The associated Fokker-Planck equation can
be immediately written down as
\begin{equation}
\frac{\partial P}{\partial t}  =  \frac{\partial}{\partial q}
\frac{V^{\prime}(q)}
{\eta}P+ \frac{k_{B}T}{\eta} \frac{\partial}{\partial q^2}P+
\Gamma \frac{\partial}{\partial q}g(q) \frac{\partial}{\partial q}g(q)P 
\end{equation}
\noindent Now, if we assume $V(q)$ and $g(q)$ to be periodic functions
with periodicity $2\pi$, the net unidirectional current can be obtained
and is given by eq.(19), with 
\begin{equation}
\psi(q)  =  \int^{q} dx \frac{V^{\prime}(x)  +  
\eta \Gamma g(x)g^{\prime}(x)}
{k_{B}T  +  \eta \Gamma g^{2}(x)},
\end{equation}
\noindent and the effective diffusion coefficient,
$$ D(q)=\frac{k_{B}T+\eta \Gamma g^{2}(q)}{\eta}.$$
For the specific form of the
periodic functions $$V(q)=V_{0}(1-cosq),$$ and \\
$$g(q)= \sqrt{g_{0}(1- \alpha cos(q- \phi))},$$ we obtain, 
$$\delta= \frac{2\pi V_{0} \sin\phi}{\eta \Gamma g_{0} \alpha}
\left[ \frac{k_{B}T+\eta \Gamma g_{0}}{\sqrt{(k_{B}T+\eta \Gamma
g_{0})^{2} - (\eta \Gamma g_{0})^{2}}} -1 \right] .$$ 
The phase $\phi$ being +ve or -ve determines the
sign of $\delta$ and consequently direction of the current $J$ (eq.(19)).
\par It should be noted that, as in case 3.2, the overdamped Brownian
particle experiences an effective space dependent temperature 
$T(q)= k_{B}T+\eta \Gamma  [g(q)]^{2}$. 
The first case (3.1) corresponds to a system which is intrinsically 
nonequilibrium and requires an internal mechanism 
such as the generation of latent heat
at the interface in first order transitions to maintain the temperature
profile $T(q)$. The other two cases (3.2 and 3.3) are, 
however, supplied with energy
externally via the externally applied white noise. And finally, we
consider a case where the Brownian particle is subjected to two 
thermal baths.

\subsection{Rectification in an inhomogeneous system under the
action of two thermal (noise) baths}

We now consider the situation in which the system is in contact
with an additive thermal noise bath at temperature $T$ and a multiplicative
thermal noise bath at temperature $\overline T$. The corresponding equation of
motion of the Brownian particle can be derived from a microscopic Hamiltonian
and is given by [22]
\begin{equation}
M{\ddot q}=-V^{\prime}(q)- \Gamma(q) \dot q + \xi_{A}(t)
+\sqrt{f(q)}\xi_{B}(t),
\end{equation}
\noindent $\xi_{A}(t)$ and $\xi_{B}(t)$ are two independent Gaussian white
noise fluctuating forces with statistics,
\begin{subequations}
\begin{eqalignno}
\langle \xi_{A}(t) \rangle\;=\;0,
\end{eqalignno}
\begin{eqalignno}
\langle \xi_{A}(t) \xi_{A}(t^{\prime})
\rangle\;=\;2\Gamma_{A}k_{B}T\delta(t-t^{\prime}),
\end{eqalignno}
\end{subequations}
\noindent and \\
\begin{subequations}
\begin{eqalignno}
\langle \xi_{B}(t) \rangle\;=\;0,
\end{eqalignno}
\begin{equation}
\langle \xi_{B}(t) \xi_{B}(t^{\prime})
\rangle\;=\;2\Gamma_{B}k_{B}{\overline T}\delta(t-t^{\prime}),
\end{equation}
\end{subequations}
\noindent where, $T$ and ${\overline T}$ are temperatures of the two baths A
and B, respectively. It should be noted that, $\xi_{A}(t)$ and
$\xi_{B}(t)$ represent internal fluctuations and together satisfy the
fluctuation-dissipation theorem $\Gamma(q)=\Gamma_{A}+\Gamma_{B}
f(q)$.  The bath B is associated with a space dependent friction
coefficient $f(q)$. When the two temperatures $T$ and $\overline T$
become equal the system will be in equilibrium and no net current can
flow.  By making $T$ and $\overline T$ different the system is
rendered nonequilibrium and one can extract energy at the expense of
increased entropy. The system, thus, acts as a Maxwell's-demon type
information engine which extracts work by rectifying internal
fluctuations.  In ref. [22] an expression for current is obtained in
the overdamped limit.  The overdamped limit of the Langevin equation
is taken by setting the left hand side of eq.(30) equal to zero. This
procedure of obtaining overdamped limit is not correct as explained in
section 1. Following the procedure of ref.[45] the correct
Fokker-Planck equation in the overdamped limit is given by [23]
\begin{eqnarray}
\frac{\partial P}{\partial t}=
 \frac{\partial}{\partial q}
\left\{ \frac{V^{\prime}(q)}{\Gamma(q)} {P}+\frac{T\Gamma_{A}}{\Gamma(q)}
\frac{\partial}{\partial q} \frac{P}{\Gamma(q)} 
+{\overline T}{\Gamma_{B}} \frac{\sqrt{f(q)}}{\Gamma(q)}
\frac{\partial}{\partial q} \frac{\sqrt{f(q)}}{\Gamma(q)} {P} \right .\nonumber\\ 
\left .+{\overline T}{\Gamma_{B}} \frac{(\sqrt{f(q)})^{\prime} \sqrt{f(q)}}
{[\Gamma(q)]^{2}} {P} \right\}.
\end{eqnarray}
\noindent For periodic functions $V(q)$ and $f(q)$ with periodicity
$2\pi$ the noise induced transport current $J$ is given by eq.(19), 
where, now
\begin{equation}
\psi(q)=\int^{q}\left\{\frac{V^{\prime}(x)\Gamma(x)}{T\Gamma_{A}+
{\overline T}{\Gamma_{B}}f(x)}+\frac{(\overline T - T)}{\Gamma(x)}
\frac{\Gamma_{A}\Gamma_{B}f^{\prime}(x)}{(T\Gamma_{A}
+{\overline T}\Gamma_{B}f(x))}\right\}dx,
\end{equation} 
\noindent and  $D(q)=\frac{T\Gamma_{A}+{\overline T}\Gamma_{B}f(q)}
{(\Gamma(q))^2}$ and
$\delta=\psi(q)-\psi(q+2\pi).$
\par As in earlier cases, taking specific periodic forms
of $V(q)=V_{0}(1- \cos q)$ and $f(q)=f_{0}(1 -\alpha \cos(q -
\phi))$, the exponent $\delta$ in eq.(19) for current is
obtained as
\begin{equation}
\delta = \left( 1 - \frac{T}{\overline T} \right)
\frac{2\pi V_{0} \sin\phi}{{\overline T} \Gamma_{B} f_{0}
\alpha} \left[ \frac{T\Gamma_{A}+{\overline T} \Gamma_{B} f_{0}}
{\sqrt{(T\Gamma_{A}+{\overline T}\Gamma_{B} f_{0})^2 -
({\overline T}\Gamma_{B} f_{0} \alpha)^2}} -1 \right]
\end{equation}  
It is clear from the expression for $\delta$ that, again as in
earlier cases (3.1-3.3), the phase difference $\phi$ between
$V(q)$ and $f(q)$ determines the direction of current $J$. It is
to be noted that, the current will flow in one direction if 
$ T >$ $\overline T$ and will flow in the opposite if $T < \overline
T$ for given $\phi$. Thus, the system acts like a Carnot engine
which extracts work by making use of two thermal baths at
different temperatures ($T \neq \overline T$). Moreover $\delta$
vanishes when $f(q)$ becomes space independent constant $f_0$,
i.e., when $\alpha = 0$, and the current $J$ becomes zero.
It should be noted further 
that when the amplitudes of $f(q)$ and $f^{\prime}(q)$ are
small compared to the amplitude of $V(q)$, the problem
turns out to be equivalent to a particle moving in a spatially varying
temperature field, $T(q)$=$(T\Gamma_{A}+{\overline
T}\Gamma_{B}f(q))/\Gamma(q)$ and, as discussed in section 3.1, such
a nonuniform temperature field yields net unidirectional current.

\section{AC driven frictional ratchets}

Having shown that the average unidirectional motion is possible in 
a periodic system subjected to an external \textbf{white noise
fluctuations}, we now turn our attention to rectification affect in
inhomogeneous systems driven by ac force as opposed to white noise
fluctuations. These ratchets are called as rocked frictional
ratchets. We study phenomena of currents, their
reversals and efficiency of energy transduction in these systems. We
obtain separately results both in adiabatic ( low frequency limit) and 
in nonadiabatic limit of external driving. We show that in adiabatic
limit, when the potential is symmetric efficiency can me maximised as
the function of noise strength or temperature. However in the case of
asymmetric potential, temperature may or may not facilitate the energy 
conversion. For this case current reversal can also be obtained in the 
adiabatic limit as the function of temperature and the amplitude of
the periodic drive. In the nonadiabatic limit
by properly tuning the parameter the system exhibits multiple current reversals, 
both as a function of thermal noise strength and as a function of
rocking force. Current reversals also occur under deterministic
conditions and exhibits intriguing structures. These results are due
to mutual interplay between potential asymmetry, noise, driving
frequency and inhomogeneous friction. 

\subsection{ Current reversals and efficiency in the adiabatic regime} 
  In this subsection we study the
motion of an over-damped Brownian particle in a potential $V(q)$ subjected
to a space dependent friction coefficient $\eta (q)$ and an external
force field $F(t)=A \cos (\omega t + \theta)$ at temperature $T$.
 The motion is described by
the Langevin equation [24,35-38]
\begin{equation}
  \label{Langv}    
  \frac{dq}{dt} = - \frac{(V'(q)-F(t))}{\eta (q)} - k_{B}T
  \frac{\eta '(q)}{[\eta (q)]^{2}} + \sqrt{\frac{k_{B}T}{\eta
      (q)}} \xi (t) ,
\end{equation}
where $\xi (t)$ is a randomly fluctuating Gaussian white noise with zero
mean and correlation ,i.e.,  $ <\xi (t) \xi (t')> = 2 \delta (t-t')$. We take
$V(q) = V_{0}(q) + qL$. For generality we take $V_{0}(q)$ to be
asymmetric periodic potential, $V_{0}(q + 2n\pi) = V_{0}(q) =
 -\sin (q) + \frac{\mu}{4} \sin(2q)$, where $\mu$ is the asymmetry
 parameter and 
$n$ being any natural integer. $L$ is a constant force ( load)
representing the slope of the washboard potential against which the
work is done. Also, we take the friction coefficient $\eta (q)$ to
be periodic : $\eta (q) = \eta _{0}(1 - \lambda \sin (q + \phi))$,
where $\phi$ is the phase difference with respect to $V_{0}(q)$. The
equation of motion is equivalently given by the Fokker-Planck equation
\begin{equation}
  \label{Fok_Plk}      
  \frac{\partial P(q,t)}{\partial t} = \frac {\partial}{\partial q}
  \frac{1}{\eta
    (q)} [k_{B}T \frac {\partial P(q,t)}{\partial q} + (V'(q) - F(t)
  )P(q,t)] .
\end{equation}
This equation can be solved for the probability current $j$ when $F(t)
= A$ = constant, and is given by [51-53]
\begin{equation}
  \label{curr}      
  j = \frac{k_{B}T ( 1 - \exp(-2 \pi(A - L)/k_{B}T))}{\int _{0}^{2
      \pi}\exp(\frac{-V_{0}(y) + (A - L)y}{k_{B}T}) dy
    \int_{y}^{y+2\pi}\eta(x)\exp \frac{V_{0}(x) - (A
      - L)x)}{k_{B}T}} dx .
\end{equation}
It may be noted that even for $L = 0,\; j(A)$ may not be equal to
$-j(-A)$  for $\phi \neq 0, \pi$. This leads to the
rectification of current (or unidirectional current) in the presence
of an applied ac field $F(t)$. We assume $F(t)$ changes slowly enough,
i.e, its frequency is smaller than any other frequency related to
relaxation rate in the problem. For a field $F(t)$ of a square wave
amplitude $A$, an average current over the period of oscillation
is given by, $<j>\; =\; \frac{1}{2}\,[j(A) + j(-A)]$. This
particle current can even flow against the applied load $L$ and
thereby store energy in a useful form. In the quasi-static limit
following the method of stochastic energetics it can be shown [54-56]
that the input energy $R$ ( per unit time) and the work $W$ ( per unit time)
that the ratchet system extracts from the external noise are given by
$R[F(t)] = \frac{1}{t_{f}-t_{i}} \int_{x=x(t_{i})}^{x=x(t_{f})} F(t)
dx(t)$ and $W = \frac{1}{t_{f}-t_{i}} \int_{x=x(t_{i})}^{x=x(t_{f})} dV[x(t)]$.
As defined earlier $V$ is the total potential including the load. The
efficiency of the energy transformation is given by $\eta =
\frac{W}{R}$. For the square wave of amplitude $A$ we get 
$R = \frac{1}{2} A[j(A)-j(-A)]$ and $W =
\frac{1}{2}L[j(A)+j(-A)]$ respectively. Thus the efficiency
( $\eta$) of the system to transform the external fluctuation
to useful work is given by
\begin{equation}
  \label{effi}         
  \eta = \frac{L[j(A) + j(-A)]}{A[j(A) - j(-A)]}.
\end{equation}
For details we refer to [54,55]. Before proceeding further we would like to emphasize that from eqn. (32), 
one can obtain the mobility of the particle in a periodic potential
tilted by a constant external field moving in a inhomogeneous
media. The mobility is given by current divided by constant external
field and is a nonlinear function of external field. Unlike the
constant friction coefficient case, where the mobility monotonically
increases and saturates to a constant value, here the mobility shows
stochastic resonance (SR) and asymptotically saturates to the previous
constant value. The occurrence of SR in the absence of weak periodic
signal is indeed a new phenomena[51-52], which essentially reduces the
constraints for the observability of SR[4,5]. Besides SR, the mobility also
shows resonance like phenomena as a function of external applied
force and noise induced stability of states ( mobility
decreases as a function of noise strength). The noise
induced stability is yet another counterintuitive novel concept aided by
the presence of noise[6,52]. This phenomena is also related to the delay of
decay of unstable states by enhancing the strength of thermal
noise.  

\begin{figure}
  \centerline{\epsfysize=14cm \epsfbox{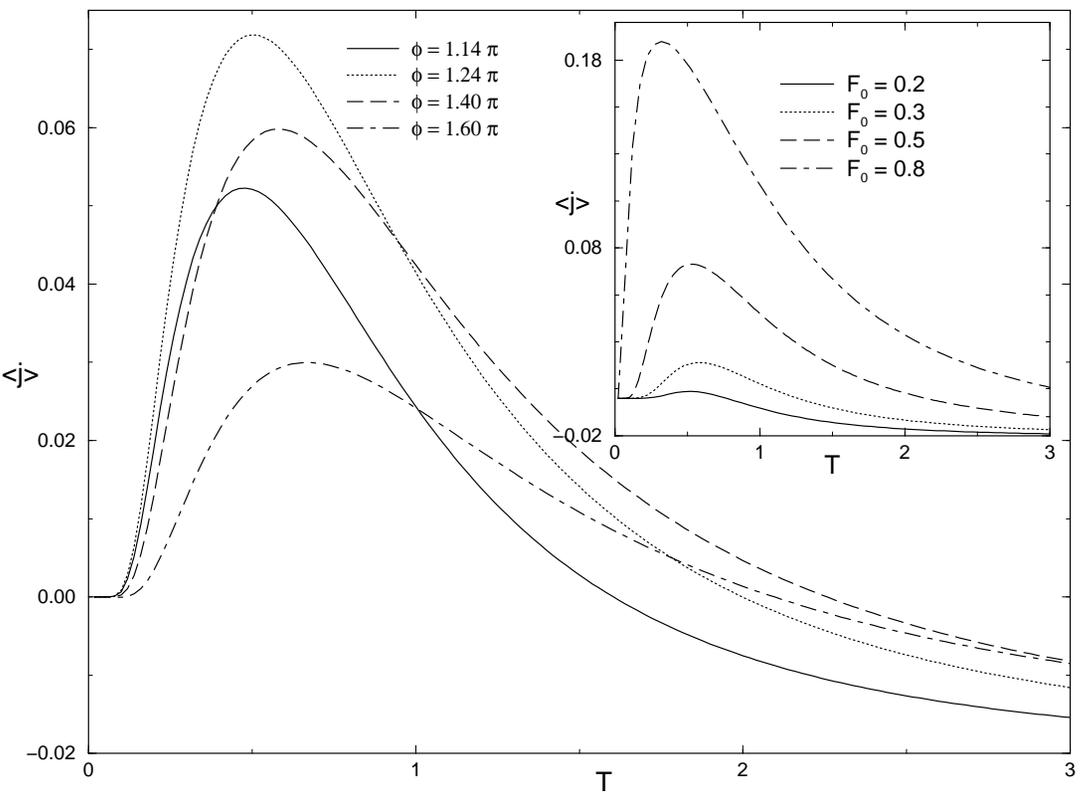}}
  \vspace{.4in}
  \caption{The net current $<j>$ as a function of $T$ for various values
of $\phi$ at $A = 0.5$. The inset shows the variation of $<j>$
with $T$ for different $A$ at $\phi = 1.3 \pi$. $L = 0.02$ for both
the figures.}
  \label{compare_curr}
\end{figure}

          We shall first discuss the effect of inhomogeneity in a
symmetric potential (i.e when $\mu = 0$) and under adiabatic
conditions and then proceed to show in the next subsection how asymmetry
and finite frequency drive brings about a qualitative change in the
current characteristics. It is found, numerically, that the effect of $\lambda$, the
amplitude of periodic modulation of $\eta(q)$, for instance on $<j>$
is more pronounced for larger value of $\lambda (0 \leq \lambda <
1)$. We choose $\lambda = 0.9$ throughout our work. Contrary to the
homogeneous friction case where the net current goes to zero , the
detailed analysis of the Eqn. (\ref{curr}) shows that the net
current $<j>$ quickly saturates monotonically to a value equal to
$\frac{-\lambda}{2} \sin \phi$ as a function of $A$ irrespective of
the value of temperature $T$. All the interesting features are
captured at smaller values of $A$ itself. We shall confine
ourselves to $A < 1$, the threshold value at which the barrier of
the potential $V_{0}(q) = -\sin q$  just disappears ( in this regime
the deterministic motion is in the locked state). In this case as
mentioned earlier it is a phase lag $\phi$ creates spatial asymmetry
responsible for currents, $\phi$ determines
the direction of asymmetry of the ratchet. For $\phi > \pi$, we have
a forward moving ratchet ( current flowing in the positive direction)
and for $\phi < \pi$ we have the opposite, in the presence of external
quasi-static force $F(t)$. When the system is homogeneous
($\lambda = 0$), $<j> = 0$ for $L = 0$ and for all values of $A$ and $T$,
because the potential $V(q)$ is symmetric. However, when $\lambda \neq
0, <j> \neq 0$ for all $\phi \neq 0, \pi$.
Fig.(\ref{compare_curr}) clearly
illustrates this. In fig.(\ref{compare_curr}) we have plotted $<j>$
(in dimensionless units) for various values of $\phi$ at
$A = 0.5 $ as a function of temperature $T$ (in dimensionless
units). Henceforth all our variables like $ <j>, A, T$ are in
dimensionless units [53].
The inset shows the behaviour of $<j>$ with temperature for
various $A$ when $\phi = 1.3 \pi$. We have chosen $L=0.02$ (corresponding to
a positive mean slope of $V(q)$). One would expect,
for $\lambda = 0$, $<j> \: < 0$ for all values of $A$ and $T \neq
0$ due to the presence of load or biasing field in the negative direction. Fig.(\ref{compare_curr}) shows that $ <j> \: > 0$ for all values
of $\phi$ ( the net current being up against the mean slope of
$V(q)$). Though the
result is counterintuitive, but it is still understandable. For
$j(-A)$ the particle is expected to have acquired higher mean
velocity before it hits the large $\eta(q)$ region and hence faces
maximum resistive force and therefore though $j(-A)$ is negative
it is small in magnitude compared to $j(A)( >0)$, where the
particle faces just the reverse situation. The average current $< j >
$ exhibits maximum as a function of temperature (SR like phenomena). At large value of
temperature, $< j >$ becomes negative. In this regime of high
temperature the ratchet effect disappears and the current which is
negative is in response to the load.
The peaking behaviour of $<j>$ as a function of $T$ is due to the
synergetic effect of the thermal
fluctuations and the space dependent friction coefficient $\eta(q)$. Of
course, this result is obtained in the quasi-static limit when the time
scale of variation of $F(t)$ is much larger compared to any other
relevant time scales involved in the system.  

\begin{figure}
  \centerline{\epsfysize=14cm \epsfbox{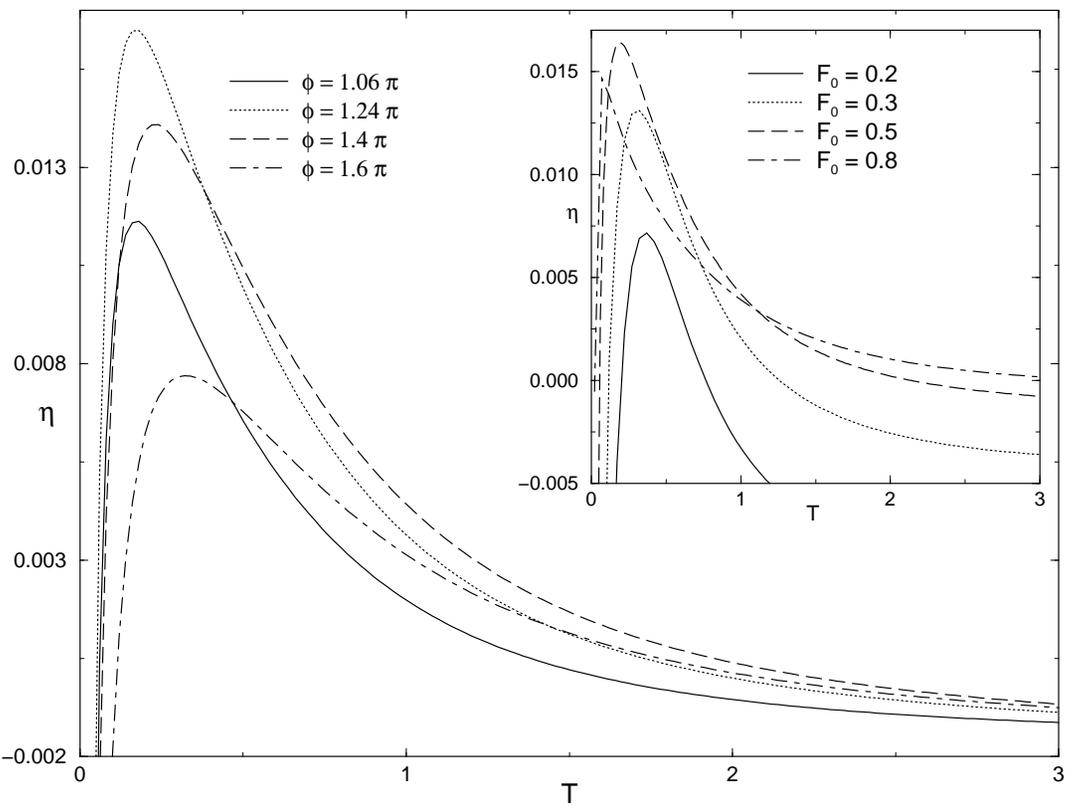}}
  \vspace{.4in}
  \caption{Efficiency $\eta$ as a function of $T$ for various values of
$\phi$ at $A = 0.5$. The inset shows the variation of $\eta$ as a
function of $T$ for different values of $A$ at $\phi = 1.3
\pi$. $L = 0.02$ for both the figures.}
  \label{compare_effi}
\end{figure}

Fig.(\ref{compare_effi}) shows the efficiency $\eta$ of flow of
current against the load (slope) $L$. The parameter values chosen for
figs.(\ref{compare_curr} and \ref{compare_effi}) are same.
\emph{$\eta$ shows a maximum as a function of temperature}.  This
implies that thermal noise can facilitate energy conversion by ratchet
system. In the adiabatically rocked ratchet this can be seen only in
the presence of space dependent friction. It is to be noted that the
temperature corresponding to maximum efficiency is not close to the
temperature at which the current $<j>$ is maximum. From this we can
conclude that \emph{the condition of maximum current does not
  correspond to maximum efficiency of current generation.} This fact
has been pointed out earlier but in a different system [55].

\begin{figure}
  \centerline{\epsfysize=14cm \epsfbox{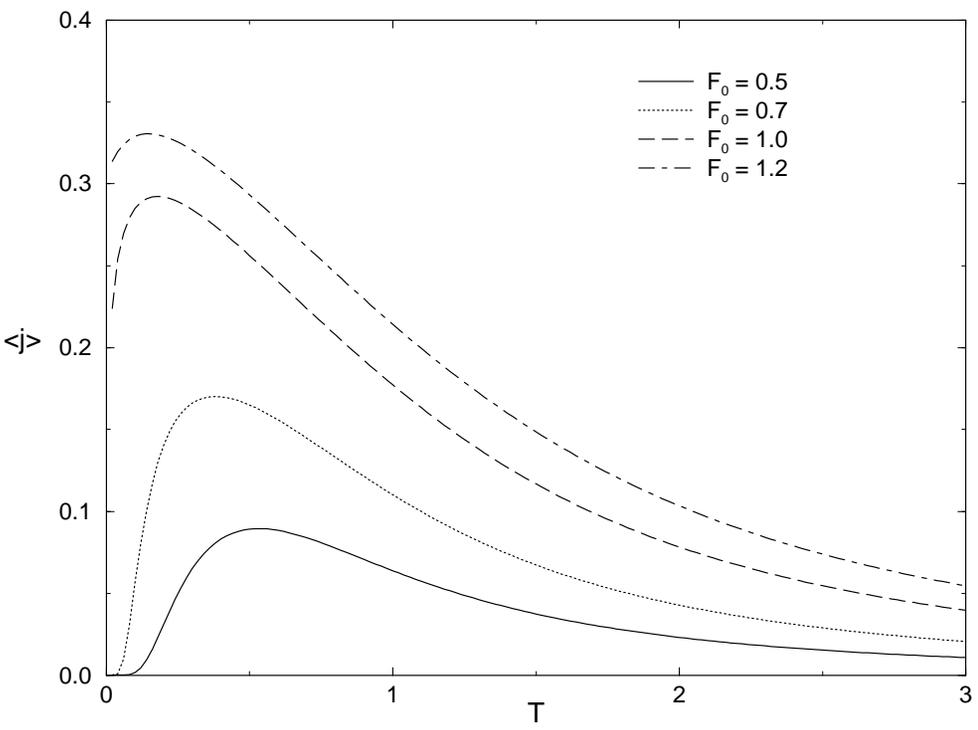}}
  \vspace{.4in}
  \caption{Current $<j>$ as a function of $T$ for various values
of $F$ at $\phi = 1.3 \pi$. $L = 0$ in this case.}
  \label{0_load}
\end{figure}

       In fig.(\ref{0_load}) we have plotted the average current $< j
>$ in the absence of load as function of temperature for various
values of $A$ for given $\phi = 1.3 \pi$. It shows that in the absence of
load $L$, $<j>$ never changes sign from positive to negative. 

\par
\begin{figure}
  \centerline{\epsfysize=14cm \epsfbox{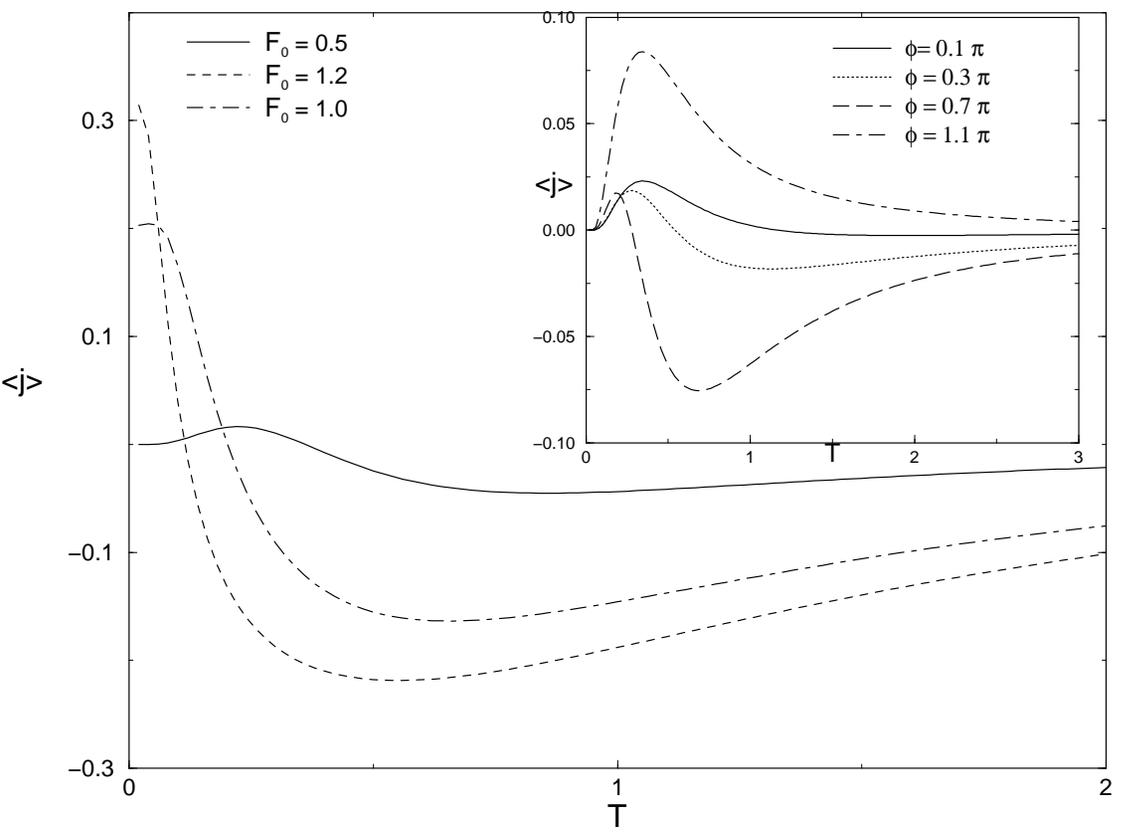}}
  \vspace{.7in}
  \caption{ Current $<j>$ in asymmetric potential for various values of
$F$ (for different values of $\phi$ in the inset) at $\phi = 0.5 \pi$( at $F =
0.5$ in the inset ) when load $L =0 $.}
  \label{curr_rev}
\end{figure}

 We now turn our attention to asymmetric potentials. It is
possible to get a current reversal as a function of
temperature if, instead of a symmetric potential one
considers an asymmetric potential $V_{0}(q) = -\sin q -\frac{\mu}{4}\sin
2q$ ( where $\mu$ lies between 0 and 1, describes an asymmetry
parameter) [57]. In fig.(\ref{curr_rev}) we have plotted $< j
>$ in the absence of load as function of temperature with asymmetry
parameter $\mu = 1 $.  In fig.(\ref{curr_rev}) we notice that the
current reverses its direction as function of temperature or noise strength.
Here too the current reversal phenomena is solely due to the space
dependent friction.  In the large temperature regime (compared to the
barrier energies) the current due to space dependent friction
dominates and hence current does not depend on the asymmetry parameter 
$\mu$. We have chosen $\phi$ such that in the high temperature limit
current is negative. Even in the absence of spatial inhomogeneity
asymmetric potential exhibits current. In our case this current is in
positive direction and dominates in the low temperature regime even
after the inclusion of inhomogeneities. Thus going from the low
temperature to the high temperature regime we have current reversal. 
Thus current reversal can be obtained by properly tuning the system
parameters.    
In homogeneous systems ( but asymmetric potential) current
reversal is possible when the frequency of the field swept $F(t)$ is
large[10,58] (nonadiabatic regime).

    In the presence of spatial asymmetry efficiency exhibits a complex 
behaviour. Depending on the system behaviour efficiency can
monotonically decrease as a function of temperature or it can show a
peaking behaviour. Thus thermal fluctuations need not facilitate the
energy conversion[36-37]. This is in contrast to the behaviour in the
presence of symmetric potential, where the efficiency always exhibits
a peaking behaviour as a function of the strength of thermal noise.

\subsection{ Current reversals in nonadiabatic regime}

            In this subsection we discuss the phenomena of current 
reversals in nonadiabatic regime in some detail. For finite frequency
drive, the probability current density $J$(x,t) is given by \\
\begin{equation}
  J(x,t) = -\frac{1}{\eta(x)} [(V'(x)-F(t)) + k_{B}T
  \frac{\partial}{\partial x}]P(x,t),
 \label{Current}
\end{equation}
where $F(t+\tau)=F(t)$ ($\tau$ being the period of the forcing term). Since the potential and the driving force
have spatial and temporal periodicity respectively, therefore $J(x,t)
= J(x + 1,t + \tau)$, [10,53]. The net current $j$
in the system is given by
$ j = \lim_{t \rightarrow \infty} \frac{1}{\tau} \int_{t}^{t+\tau} dt
 \int_{0}^{1} J(x,t) dx$.
  $j$ is
independent of the initial phase $\theta$ of the driving force. We
solve the FPE numerically by the method of finite difference and calculate
the current $j$[38].

\begin{figure}
\protect\centerline{\epsfysize=2.7in \epsfbox{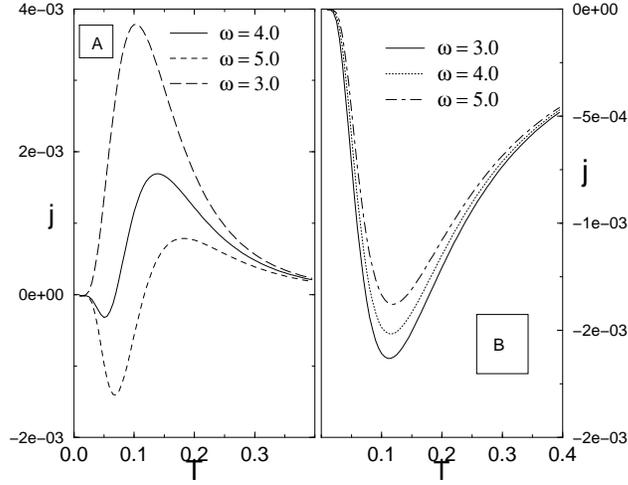}}
\caption{Mean current j vs temperature T for (A) $\lambda = 0, \mu =
  1.0$ and (B) $\lambda = 0.1, \mu = 0$ and $\phi = 0.2\pi$. Note
  there is no current reversal when the potential is symmetric.}
\label{lam=0}
\end{figure}

                    In the Fig.~(\ref{lam=0}A), the average current $j$
is plotted as a function of temperature $T$ for different values of
$\omega$. Here the asymmetry parameter $\mu =
1.0$ and $\lambda = 0.0$.
       In this case the current reverses its sign
(only \textit{once}) for frequencies sufficiently
large as shown in the $\omega = 4.0, \omega = 5.0$ case. These
frequencies corresponds to nonadiabatic regime [10] and are larger than
intrawell relaxation frequency $\omega_{0}=3.18$. Current in the
adiabatic limit can be obtained analytically[36-37]. In the
absence of asymmetric potential and presence of space dependent
friction ( $\lambda = 0.9$), there is no current reversal irrespective
of $\omega$ and phase shift $\phi$ as shown in Fig~(\ref{lam=0}B). Hence asymmetric potential
is essential for current reversal. However, the direction of the current
depends on the phase lag $\phi$. Separately in both these cases
 absolute value of current
 exhibits a maxima as a function of $T$, reminiscent of SR.  In the
 absence of space dependent friction fig. (6A), for
frequencies higher than intrawell frequency $\omega_{0}$ , the low
temperature scenario is governed by  
interplay between potential asymmetry and $\omega$. Due to higher
frequency the Brownian particles do not get 
enough time to cross the barriers. Most of the particles move about
the potential minima. The probability of finding the
Brownian particles near the minima increases with
increasing frequency, consequently the probable number of particles
near the potential barrier decreases. Since the distance from a
potential minima to the basin of attraction of next minima is less
from steeper side (which is at left side of the minima) than from the
slanted side, hence in one period 
the particles get enough time to climb the potential barrier from the
steeper side than from the slanted side, resulting in a negative
current. On increasing  the temperature, the particles get kick of
larger intensity and hence they easily cross the slanted barrier,
resulting in a current reversal and positive current. 

\begin{figure}
  \protect\centerline{\epsfysize=2.5in \epsfbox{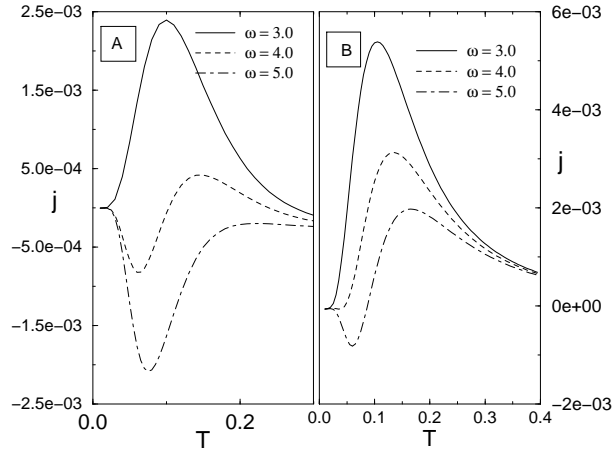}}
  \caption{ The mean current $j$ \textit{vs} temperature $T$ for $\phi =
    0.2 \pi$, $A=0.5$ and $\lambda = 0.1$. The driving frequencies are $\omega=
    3.0, 4.0$ and $5.0$. The right hand side figure shows  current $j$
    \textit{vs} $T$ for $\phi = 1.2 \pi$.}
  \label{j-T}
\end{figure}

            Now we study the combined effect of spatial asymmetry and
frictional inhomogeneity in the nonadiabatic regime. In fig. (7A)
we have plotted $j$ \textit{vs} $T$ with $\mu = 1.0$, $\lambda = 0.1$
and $\phi = 0.2 \pi$ for different values of $\omega$.
In the presence
of finite frequency drive there are two current reversals as shown
in the figure (for the case of $\omega = 4.0$). This phenomena of
twice current reversal with 
 temperature $T$ is the foremost feature in our model, previously
 unseen in any overdamped system. In this case $\phi$ has been chosen 
 in such a manner so that current in the high temperature limit (
 which is predominantly determined by frictional inhomogeneity than
 spatial asymmetry) is opposite to pure spatially asymmetric case
 ($\lambda=0.0$). This is the cause of second current reversal as seen
 for the $\omega = 4.0 $ case. The first current reversal being the
 effect of finite frequency driving and spatial asymmetry as discussed 
 earlier. When 
the phase difference $\phi$ is such that the current due to space
dependent friction alone is in the same direction as that of current due to
potential asymmetry only, then we do not have current reversal in the
adiabatic case as shown in Fig~(\ref{j-T}B) where $\phi = 1.2 \pi$,
though a \textit{single} current reversal due to
finite frequency drive may be present. In all the cases studied so far
we have observed that the current reversal do not take place above a
critical frequency $\omega_{c}$ of driving, which in turn depends
sensitively on $\phi$ and other parameters in the problem.

\begin{figure}
 \protect\centerline{\epsfysize=3.0in \epsfbox{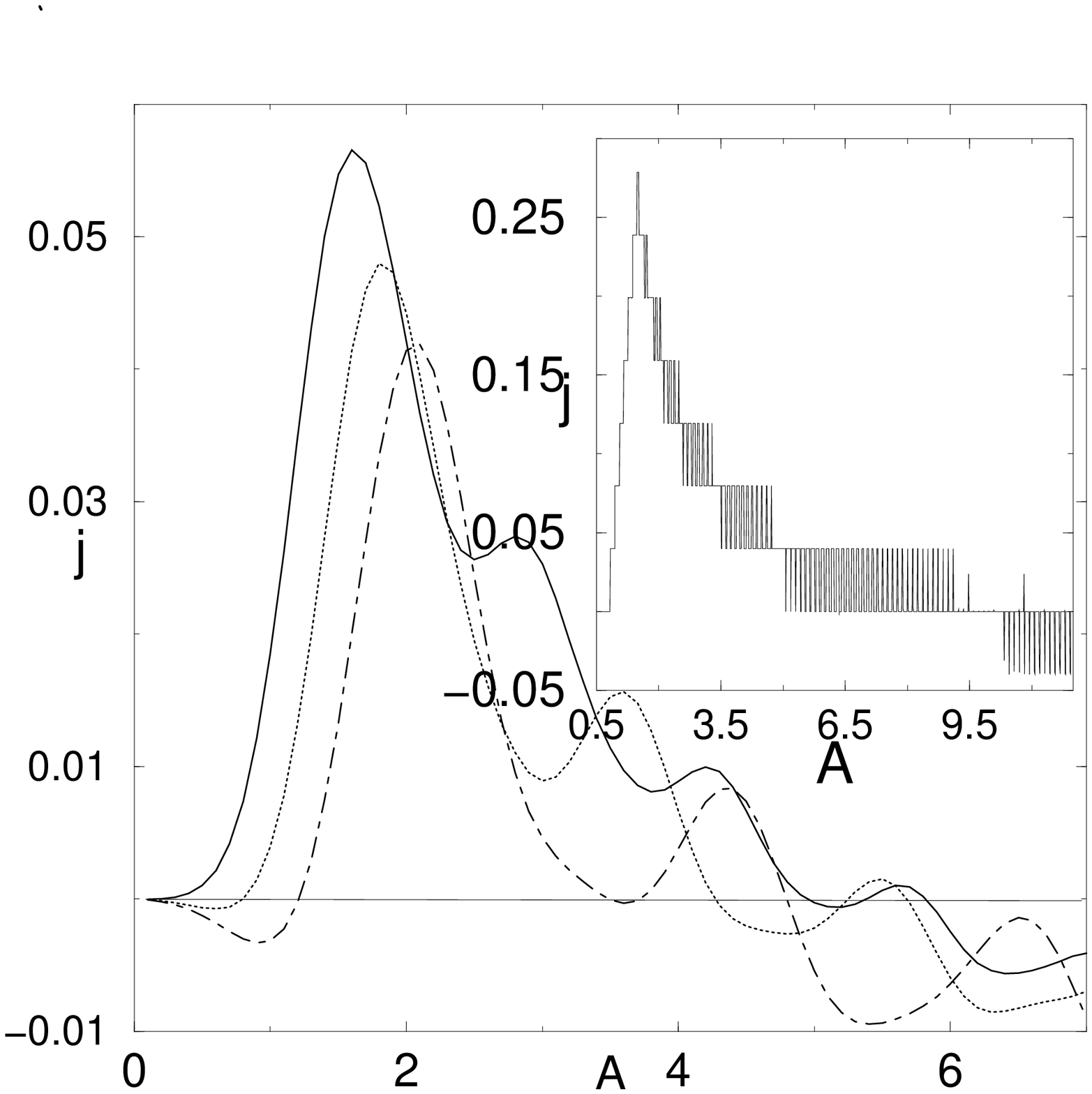}}
\caption{The mean current $j$ with amplitude $A$ of the forcing term
  for $\phi = 0.88 \pi,T=0.05$ and $\lambda=0.1$ with $\omega=3.0,
  4.0, 5.0$. The inset shows the reversal of deterministic current vs the
  amplitude of the forcing for $\omega=0.25$.}
\label{j-A}
\end{figure}

Multiple current reversals can also be seen when the amplitude ( $A$) of the
forcing term is varied in a suitable parameter regime of our system.
In Fig.~\ref{j-A}, the plot of $j$ versus $A$ is shown for
different values of $\omega$, keeping $\lambda$, $\phi$ and $T$ fixed at
$0.1$, $0.88\pi$ and $0.05$ respectively. For $\omega = 4.0$ curve, we can see as
many as four current reversals. For very large value of $A$, the current
asymptotically goes to a constant value $-\frac{\lambda}{2}\sin
(\phi)$,  as was previously shown for the adiabatic
case. It should be noted that in the same asymptotic
regime current goes to zero in the absence of space dependent friction. As 
the asymptotic value of current depends on $\phi$, so we can choose it
appropriately to make it positive or negative. In
the present case, $\phi$ has been chosen such that the asymptotic
current is negative which guarantees at least one current reversal
irrespective of frequency. The oscillatory behaviour in the $j-A$
characteristics is the reminiscent of deterministic dynamics. The inset in
Fig.~\ref{j-A} shows current reversal even for the deterministic case also.
 In addition it exhibits the interesting phenomena of current
quantization and phase locking. This current quantization give rise to 
oscillatory behaviour due to broadening of the steps in presence of
small thermal noise. On further increasing the temperature these oscillatory 
behaviour vanishes. 

\begin{figure}
\protect\centerline{\epsfysize=2.7in \epsfbox{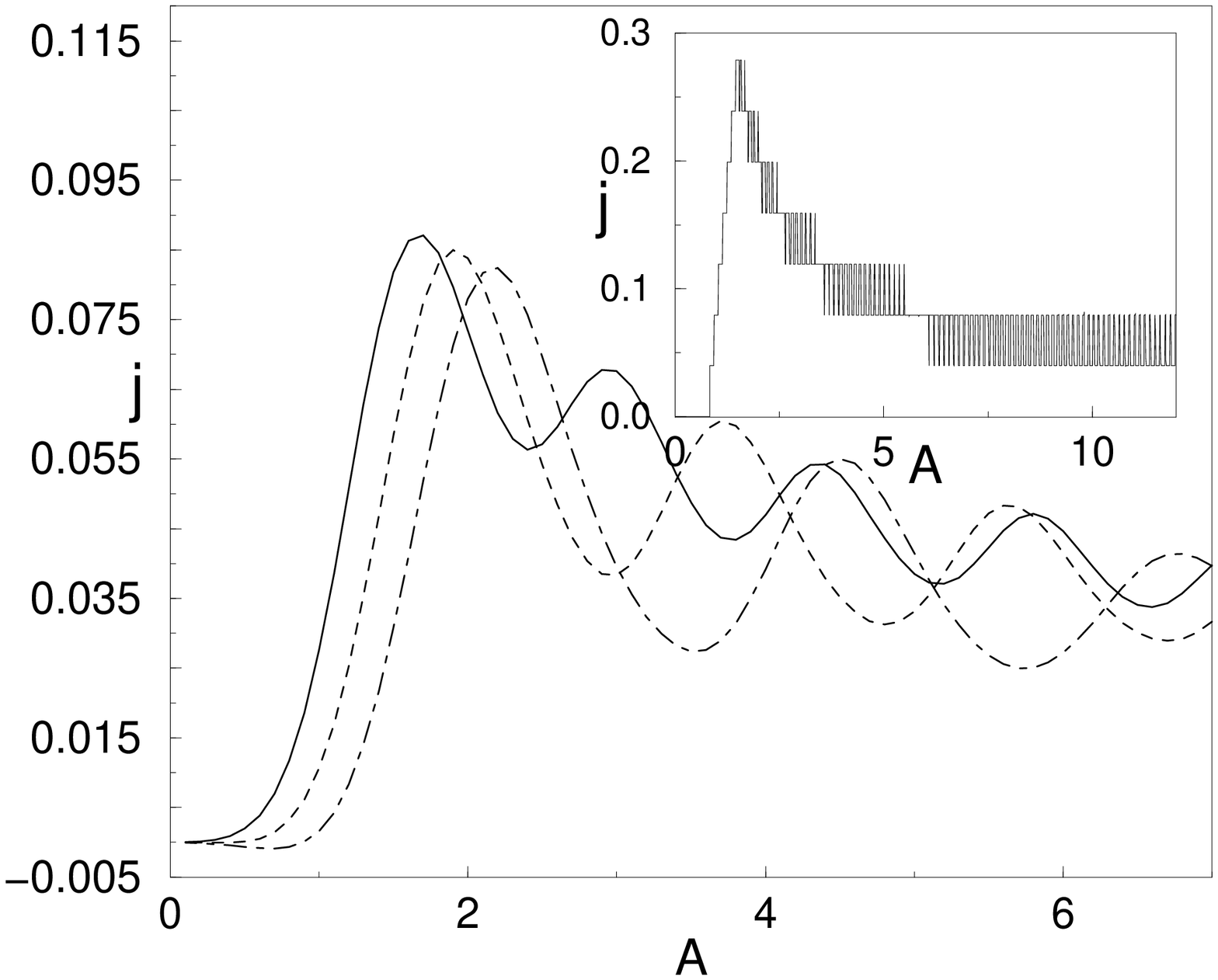}}
\caption {The mean current $j$ with amplitude $A$ of the forcing term
  for $\phi = 1.2 \pi,T=0.05$ and $\lambda=0.1$ with $\omega=3.0,
  4.0, 5.0$. The inset shows the deterministic current vs the
  amplitude of the forcing for $\omega=0.25$.}
\label{j-A2}
\end{figure}

In Fig~(\ref{j-A2}) we have plotted  $j$ versus $A$ for $\phi =
1.2 \pi$ for various values of $\omega$. There is no current reversal
in the adiabatic regime. Here $\phi$ has been chosen such that, the
asymptotic current is in the same direction as that due to the spatial 
asymmetry of the potential (i.e positive). Hence no multiple current
reversal can be seen in the nonadiabatic regime even though there is
oscillatory behaviour. The only reversal at
low value of $A$ is that of finite frequency drive as discussed
previously.  The observation 
of multiple current reversals can be attributed to a cooperative interplay
between the spatial asymmetry of the potential, the friction
inhomogeneity and the finite frequency drive. Depending on the system
parameters we may have multiple current reversal or no current
reversal at all ( see Fig.(~\ref{j-A} and \ref{j-A2})).
All the above results can be understood qualitatively.

\section{Summary and Discussion} 

\par Transport in a nonequilibrium periodic system has become, in
recent times, a field of very active research. We have just
tried, in the beginning of this work, only to enumerate various
working ideas to build a plausible model of thermal
ratchet. The brief enumeration is, of course, not complete. The
models are being gradually refined and simplified to be close
either to the experimental reality or to invent techniques
to be useful in practice. 
Transport in inhomogeneous nonequilibrium systems has attracted
attention since long. We have presented in Sec.2  a microscopic approach
to obtain macroscopic equation of motion in such systems. 

\par With the help of these equations in Sec. 3.1  we have considered
a medium with space dependent temperature and friction and have
discussed the origin of unidirectional current in these systems. In
the situations considered,  in sections 3.1 and 3.2, 
the system is subjected to external white noise fluctuations violating
the fluctuation-dissipation theorem.
Also, these two cases, in a sense, are
physically equivalent to having a spatially varying temperature 
field as considered in section 3.1. In 
section 3.4 we have considered an inhomogeneous system under the action
two thermal baths. This system acts like a Carnot engine which
extracts work by making use of two thermal baths being at different
temperature. This model too in the limit
of small friction field modulation amplitude 
corresponds to a spatially varying temperature field. These
observations seem to suggest that the case of inhomogeneous
systems with spatially varying temperature field provides a
general paradigm to obtain macroscopic current and several
variants considered to model fluctuation induced transport may
fall in the same general class of problems as considered in section
3.1. Moreover in all our models to obtain unidirectional current we
require external white noise 
fluctuations as opposed to various models for homogeneous systems
where application of colored noise (correlated noise) is must.  
For example, earlier models of thermal ratchets driven by colored noise
in a small correlation time expansion (or in the unified colored noise
approximation for arbitrary time) become identical 
to a Brownian particle moving in an
inhomogeneous medium with space dependent diffusion coefficient [59].
The interesting idea of relative stability of states,
which affects current, in the presence of temperature 
nonuniformity which is central to our treatment, 
however, has not received the attention it deserves in the area of
nonequilibrium thermodynamics. 
We remark that we do not require the periodic
potential field of the system to be ratchetlike nor do we
require the fluctuating force to be a colored noise to obtain
macroscopic current.
Thus we have put the
problem of macroscopic unidirectional motion in
nonequilibrium systems on a more general footing.

                          In Sec. 4 we have discussed the current
rectification and their reversals in ac driven frictional ratchet both 
in adiabatic and in nonadiabatic regime. Both spatially symmetric and
asymmetric potential has been considered. We observe several novel
and complex features arising due to the combined effect of asymmetry and
inhomogeneous friction and finite frequency driving force.
Currents in the low temperature regime is mostly influenced by the asymmetry
of the potential. At higher temperatures it is controlled by the modulation
parameter $\lambda$ of the friction coefficient. We find current reversal
with temperature even when the forcing is adiabatic. In the presence of
finite frequency, twice current reversal occurs. As function of
amplitude of the forcing term we observe multiple current reversals.
Current even
reverses its sign in the adiabatic deterministic regime and exhibits
intriguing feature like phase locking and current quantization.
Current reversals in ratchets are very sensitive to the
nature of potential and system
parameters. Even the condition of current direction cannot be readily
predicted a priori. Ratchets of different kind (and there are as many
as discussed by Reimann [8]) require different conditions
for current reversals, which have been worked out in some limiting
cases [8]. However, all ratchets exhibit great sensitivity
to the form of underlying potential and its derivatives over the
entire period (see the introduction of [8]). For example,
it has recently been shown [60] that by barrier subdivision
of potential profile with an integer n number of modulations (in a
randomly rocked ratchet) can lead to n-1 current reversals as a
function of temperature. However, there is no universal rule to obtain
current reversal.  
We expect that our analysis should be applicable for the motion of particle
in porous media and for molecular motors where space dependent friction can
arise due to the confinement of particles.

         We have briefly discussed the efficiency of energy conversion 
in these systems in the adiabatic limit. We have shown that efficiency 
for energy transduction can be maximised as a function of noise
strength and temperature. This in itself is a interesting
observation. However, in the case of asymmetric potential the
temperature may or may not facilitate the energy conversion. Some of
our recent studies reveal that in nonadiabatic limit, efficiency of
the system may be small or large compared to the adiabatic regime. 
This indicates that by going away from the quasi-static
limit(adiabatic limit) efficiency  can be increased contrary to  the
behaviour for reversible engines.  For heat engines at molecular
levels(molecular motors) the generalisation  of thermodynamic
principles to nonequilibrium steady states is a subject of  current
interest[61]. In a recent work[32] it has been  shown that in two
state ratchet models(flashing ratchets)  mobility can take over the
role of the potential. Contrary to previous models even no microscopic
forces(or potentials) are involved in the transport mechanism.  We
further expect that several novel cooperative effects and quantum
effects discussed in the introduction for homogeneous systems may lead
to a rich variety of phenomenon in inhomogeneous systems. System
inhomogeneities may further enhance our  understanding of  non-Debye
relaxation (approach to equilibrium) even in the absence of potential
disorder.  To conclude we expect that the further studies of dynamics
of particles in inhomogeneous non equilibrium systems may lead to new
concepts in thermodynamics and statistical physics. 

\section{Acknowledgement}

          The author thanks Debasis Dan, Dr. T .P .Pareek  and Dr. M. C. Mahato
for several useful discussions on this subject. The author also acknowledges
 Sandeep K. Joshi and Colin Benjamin for their help in preparaing this
manuscript.


\end{document}